\documentclass{IEEEtran}
\usepackage{amsmath,amssymb,url}
\newtheorem{theorem}{Theorem}
\newtheorem{lemma}{Lemma}
\newtheorem{definition}{Definition}

\newtheorem{example}{Example}

\usepackage{subfigure}
\usepackage{epsfig}

\def\f{{\mathbb{F}}}

\providecommand{\abs}[1]{\lvert#1\rvert}
\providecommand{\norm}[1]{\lVert#1\rVert}

\begin{document}

\title{Square Complex Orthogonal Designs with no Zero Entry for any $2^m$ Antennas}
\author{Smarajit Das,~\IEEEmembership{Student Member,~IEEE} and
        B. Sundar Rajan,~\IEEEmembership{Senior Member,~IEEE}%
\thanks{This work was supported through grants to B.S.~Rajan; partly by the
IISc-DRDO program on Advanced Research in Mathematical Engineering, and partly
by the Council of Scientific \& Industrial Research (CSIR, India) Research
Grant (22(0365)/04/EMR-II).} 
\thanks{Smarajit Das and B. Sundar Rajan are with the Department of Electrical Communication Engineering, Indian Institute of Science, Bangalore-560012, India.email:\{smarajit,bsrajan\}@ece.iisc.ernet.in.}}

\maketitle

\begin{abstract}
Space-time block codes from square complex orthogonal designs (SCOD) have been extensively studied and most of the existing SCODs contain large number of zeros. The zeros in the designs result in high peak-to-average power ratio and also impose a severe constraint on hardware implementation of the code while turning off some of the transmitting antennas whenever a zero is transmitted. 
Recently, SCODs with no zero entry have been constructed for $2^a$ transmit antennas whenever $a+1$ is a power of $2$. Though there exists codes for $4$ and $16$ transmit antennas with no zero entry, there is no general method of construction which gives codes for any number of transmit antennas. 
In this paper, we construct SCODs for any power of $2$ number of transmit antennas having all its entries non-zero.
Simulation results show that the codes constructed in this paper outperform the existing codes for the same number of antennas under peak power constraint while performing the same under average power constraint.
\end{abstract}

\section{Preliminaries}
Space-Time Block Codes (STBCs) from complex orthogonal designs (CODs) have been extensively studied in \cite{TJC,TiH,Lia}. 
Due to the orthogonality of the designs, the codes have linear decoding complexity, that is, they are single symbol decodable (SSD). Generally,
a {\textit{linear-processing complex orthogonal design}} (LPCOD) is a $p\times n$ matrix $G$ in $k$ complex variables $x_1,x_2,\cdots,x_k$ such that each non-zero entry of the matrix is a complex linear combinations of the variables $x_1,x_2,\cdots,x_k$ and their conjugates $x_1^*,x_2^*,\cdots,x_k^*$ satisfying $G^\mathcal{H}G=({\vert x_1\vert}^2 +{\vert x_2\vert}^2+\cdots+{\vert x_k\vert}^2)I_n$, where $G^\mathcal{H}$ is the complex conjugate transpose of $G$ and $I_n$ is the $n\times n$ identity matrix. 
An LPCOD $G$ is called {\textit{complex orthogonal design}} (COD) if the non-zero entries of $G$ are the complex variables $\pm x_1,\pm x_2,\cdots,\pm x_k$ or their complex conjugates (entries with complex linear combinations of the variables and their conjugates are not allowed).

For the construction of codes with low peak-to-average power ratio (PAPR), we relax the conditions imposed on the entries of a COD. We define
 {\textit {$\lambda$-scaled square complex orthogonal design}}, for a positive integer $\lambda$, ($\lambda$-scaled COD) $G$ as a $n\times n$ matrix in $k$ complex variables $x_1,x_2,\cdots,x_{k}$ such that any non-zero entry of the matrix is a variable or its complex conjugate, or the negative of these and all the entries of any subset of columns of the matrix is scaled by $\frac{1}{\sqrt{\lambda}}$ satisfying the condition: $G^\mathcal{H} G=({\vert x_1\vert}^2 +\cdots+{\vert x_{k}\vert}^2)I_n$.
Notice that a $\lambda$-scaled COD with no column scaled by $\frac{1}{\sqrt{\lambda}}$ is a COD (corresponds to $\lambda =1$). In columns with scaling by $\frac{1}{\sqrt{\lambda}}$ all the variables appear exactly $\lambda$ times. 
In this paper, $\lambda$ is always a power of $2$ and call these codes simply {\textit{scaled-COD}}s. 
To construct codes with all its entries non-zero, the notion of co-ordinate interleaved complex variables is found to be useful. This type of variable is used extensively in the  construction of single-symbol decodable STBCs that are not CODs \cite{KhR}.
Given two complex variables $x_i$ and $x_k$ where $x_i=x_{iI}+jx_{iQ}$ and $x_k=x_{kI}+jx_{kQ}$, the coordinate interleaved variables corresponding to the variables $x_i$ and $x_k$, are $x_{i,k}=x_{iI}+jx_{kQ}$ and $x_{k,i}=x_{kI}+jx_{iQ}.$ 
\begin{definition}
An LPCOD is called {\textit {coordinate interleaved scaled complex orthogonal designs}} (CIS-COD) if any non-zero entry of the matrix is a variable or a coordinate interleaved variable, or their complex conjugates, or multiple of these by $\pm \frac{1}{\sqrt{\lambda}}$ where $\lambda$ is a power of $2$.
\end{definition}
Note that any scaled-COD is a CIS-COD, but not conversely.

\begin{figure*}
{\small
\begin{equation*}
\label{TWMS8code}
G_{TWMS}=
\frac{1}{\sqrt{2}}\left[
\begin{array}{r@{\hspace{0.9pt}}r@{\hspace{0.9pt}}r@{\hspace{0.9pt}}r@{\hspace{0.9pt}}r@{\hspace{0.9pt}}r@{\hspace{0.9pt}}r@{\hspace{0.9pt}}r@{\hspace{0.9pt}}}
x_{1} &x_{1} &x_{2} &x_{2} &x_{3} &x_{4} &x_{3} &x_{4}  \\
x_{1} &-x_{1} &x_{2} &-x_{2} &x_{4}^* &-x_{3}^* &x_{4}^* &-x_{3}^*  \\
x_{2}^* &x_{2}^* &-x_{1}^* &-x_{1}^* &x_{3} &x_{4} &-x_{3} &-x_{4}  \\
x_{2}^* &-x_{2}^* &-x_{1}^* &x_{1}^* &x_{4}^* &-x_{3}^* &-x_{4}^* &x_{3}^*  \\
x_{4I}+jx_{3Q}  &x_{3I}+jx_{4Q} &x_{4I}+jx_{3Q} &x_{3I}+jx_{4Q} &x_{2I}+jx_{1Q} &x_{2I}+jx_{1Q} &x_{1I}+jx_{2Q} &x_{1I}+jx_{2Q}  \\
x_{3I}+jx_{4Q}  &x_{4I}+jx_{3Q} &x_{3I}+jx_{4Q} &x_{4I}+jx_{3Q} &x_{2I}+jx_{1Q} &x_{2I}+jx_{1Q} &x_{1I}+jx_{2Q} &x_{1I}+jx_{2Q}  \\
x_{4I}+jx_{3Q}  &x_{3I}+jx_{4Q} &x_{4I}+jx_{3Q} &x_{3I}+jx_{4Q} &x_{1I}+jx_{2Q} &x_{1I}+jx_{2Q} &x_{2I}+jx_{1Q} &x_{2I}+jx_{1Q}  \\
x_{3I}+jx_{4Q}  &x_{4I}+jx_{3Q} &x_{3I}+jx_{4Q} &x_{4I}+jx_{3Q} &x_{1I}+jx_{2Q} &x_{1I}+jx_{2Q} &x_{2I}+jx_{1Q} &x_{2I}+jx_{1Q}
\end{array}
\right]
\end{equation*}
}
\hrule
\end{figure*}
It is known that the maximum rate $\mathcal{R}$ of an $n\times n$ LPCOD is $\frac{a+1}{n}$ where $n=2^a(2b+1), a\mbox{ and } b$ are positive integers \cite{TiH}.
Several authors have constructed LPCODs for $2^a$ antennas achieving maximal rate \cite{TiH,ALP,Joz,Wol}. In \cite{TiH}, the following induction method is used to construct SCODs for $2^a$ antennas, $a=2,3,\cdots$, starting from {\small  $G_1= \left[ 
\begin{array}{ c @{\hspace{.2pt}} c @{\hspace{.2pt}}}
x_1   &-x_2^*      \\
x_2   & x_1^*
\end{array}\right],$}
\begin{equation}
\label{itcod}
G_a=  \left[ 
\begin{array}{ c @{\hspace{.4pt}} c @{\hspace{.2pt}}}
G_{a-1}   & -x_{a+1}^*I_{2^{a-1}}      \\
x_{a+1}I_{2^{a-1}}   & G_{a-1}^{\mathcal{H}}
\end{array}\right]
\end{equation} 
\noindent
where $G_a$ is a $2^a\times 2^a$ complex matrix. Note that $G_a$ is a
COD in $a+1$ complex variables $x_1,x_2,\cdots, x_{a+1}$. Moreover, each row and each column of the matrix $G_a$ contains only $a+1$ non-zero elements and all other entries in the same row or column are filled with zeros. The fraction of zeros, defined as the ratio of the number of zeros to the total number of entries in a design, for $G_a$, is
\begin{equation}
\label{largefrac}
\frac{2^a-a-1}{2^a}=1-\frac{a+1}{2^a}=1-\mathcal{R}.
\end{equation}
For the constructions in \cite{TiH,ALP,Joz,Wol} also, the fraction of zeros is given by \eqref{largefrac}. Reducing number of zeros in a SCOD for more than $2$ transmit antennas (for two antennas, the Alamouti code does not have any zeros), is important for many reasons including improvement in Peak-to-Average Power Ratio (PAPR) and also the ease of practical implementation of these codes in wireless communication system \cite{YGT1}.

For $8$ transmit antennas, the SCOD $G_3$ obtained by the construction \eqref{itcod} as shown below 
\begin{eqnarray}
G_3=\left[\begin{array}{r@{\hspace{0.9pt}}r@{\hspace{0.9pt}}r@{\hspace{0.9pt}}r@{\hspace{0.9pt}}r@{\hspace{0.9pt}}r@{\hspace{0.9pt}}r@{\hspace{0.9pt}}r@{\hspace{0.9pt}}}
    x_1   &-x_2^*  &-x_3^*& 0    &-x_4^* & 0     & 0    & 0 \\
    x_2   & x_1^*  & 0    &-x_3^*& 0     &-x_4^* & 0    & 0 \\
    x_3   & 0      & x_1^*& x_2^*& 0     & 0     &-x_4^*& 0 \\
     0    & x_3    &-x_2  & x_1  & 0     & 0     & 0    &-x_4^* \\
    x_4   & 0      & 0    & 0    & x_1^* & x_2^* & x_3^*& 0 \\
     0    & x_4    & 0    & 0    &-x_2   & x_1   & 0    & x_3^*\\
     0    & 0      & x_4  & 0    &-x_3   & 0     & x_1  &-x_2^*\\
     0    & 0      & 0    & x_4  & 0     &-x_3   & x_2  & x_1^*\\
\end{array}\right],\\
\end{eqnarray}

\noindent
contains $50$ per cent of entries zeros. But, Yuen et al, in \cite{YGT}, have constructed a new rate-$1/2$, SCOD $\frac{G_Y}{\sqrt{2}}$ of size $8$ with no zeros in the design matrix using Amicable Complex Orthogonal Design (ACOD) \cite{GeS} where $G_Y$ is given in \eqref{itcod8}.
\begin{eqnarray}
\label{itcod8}
G_Y=\left[\begin{array}
{r@{\hspace{0.4pt}}r@{\hspace{0.4pt}}r@{\hspace{0.4pt}}r@{\hspace{0.4pt}}r@{\hspace{0.4pt}}r@{\hspace{0.4pt}}r@{\hspace{0.4pt}}r@{\hspace{0.4pt}}}
x_{1}^*  & x_{1}^*& x_{2}   &-x_{2}   &x_{3}   &-x_{3}  & x_{4}  & -x_{4}   \\
jx_{1}   & -jx_{1}  & jx_{2}^* &jx_{2}^* &jx_{3}^* & jx_{3}^*& jx_{4}^*& jx_{4}^* \\
-x_{2}   & x_{2}  & x_{1}^* &x_{1}^* &x_{4}^* & -x_{4}^*& -x_{3}^*& x_{3}^* \\
-jx_{2}^*& -jx_{2}^*& jx_{1}   &-jx_{1}   &jx_{4}   & jx_{4}  & -jx_{3}  & -jx_{3}   \\
-x_{3}   & x_{3}  & -x_{4}^* &x_{4}^* &x_{1}^* & x_{1}^*& x_{2}^*& -x_{2}^*   \\
-jx_{3}^*& -jx_{3}^*& -jx_{4}   &-jx_{4}   &jx_{1}   & -jx_{1}  & jx_{2}  & jx_{2}   \\
-x_{4}   & x_{4}  & x_{3}^* &-x_{3}^* &-x_{2}^* & x_{2}^*& x_{1}^*& x_{1}^*   \\
-jx_{4}^*& -jx_{4}^* & jx_{3}  &jx_{3}   &-jx_{2}   & -jx_{2}  & jx_{1}  & -jx_{1}
\end{array}\right]
\end{eqnarray}
\begin{figure*}
\begin{equation}
\label{NZE1}
\frac{1}{\sqrt{2}}\left[\begin{array} 
{r@{\hspace{0.9pt}}r@{\hspace{0.9pt}}r@{\hspace{0.9pt}}r@{\hspace{0.9pt}}r@{\hspace{0.9pt}}r@{\hspace{0.9pt}}r@{\hspace{0.9pt}}r@{\hspace{0.9pt}}r@{\hspace{0.9pt}}r@{\hspace{0.9pt}}r@{\hspace{0.9pt}}r@{\hspace{0.9pt}}r@{\hspace{0.9pt}}r@{\hspace{0.9pt}}r@{\hspace{0.9pt}}r@{\hspace{0.9pt}}}
x^*_1&x^*_1&x^*_2&x^*_2&x_3&-x_3&x^*_4&x^*_4 & x_5/2&x_5/2&x_5/2 & x_5/2 &x_5/2&x_5/2&x_5/2&x_5/2\\
x_1&-x_1&x_2&-x_2&x^*_3&x^*_3&x_4&-x_4 &x_5/2&-x_5/2&x_5/2 & -x_5/2&x_5/2&-x_5/2&x_5/2&-x_5/2\\
-x_2&-x_2&x_1&x_1&-x^*_4&-x^*_4&x_3&-x_3 & x_5/2&x_5/2&-x_5/2 & -x_5/2&x_5/2&x_5/2&-x_5/2&-x_5/2\\
-x^*_2&x^*_2&x^*_1&-x^*_1&-x_4&x_4&x^*_3&x^*_3 & x_5/2&-x_5/2&-x_5/2 & x_5/2&x_5/2&-x_5/2&-x_5/2&x_5/2\\
-x_3&x_3&x_4&x_4&x^*_1&x^*_1&-x_2&-x_2  &x_5/2&x_5/2&x_5/2 & x_5/2&-x_5/2&-x_5/2&-x_5/2&-x_5/2\\
-x^*_3&-x^*_3&x^*_4&-x^*_4&x_1&-x_1&-x^*_2&x^*_2  &x_5/2&-x_5/2&x_5/2 & -x_5/2&-x_5/2&x_5/2&-x_5/2&x_5/2\\
-x_4&-x_4&-x_3&x_3&x^*_2&x^*_2&x_1&x_1  & x_5/2&x_5/2&-x_5/2 & -x_5/2&-x_5/2&-x_5/2&x_5/2&x_5/2\\
-x^*_4&x^*_4&-x^*_3&-x^*_3&x_2&-x_2&x^*_1&-x^*_1  & x_5/2&-x_5/2&-x_5/2 & x_5/2&-x_5/2&x_5/2&x_5/2&-x_5/2\\
 -x^*_5/2 & -x^*_5/2 & -x^*_5/2 & -x^*_5/2 & -x^*_5/2&-x^*_5/2&-x^*_5/2&-x^*_5/2  & 
x_{1,2}&x_{1,2}&x_{4,1} &x_{4,1}&x_3&-x_3&x_{2,4}&x_{2,4}\\
-x^*_5/2 & x^*_5/2 & -x^*_5/2 & x^*_5/2 & -x^*_5/2 & x^*_5/2& -x^*_5/2&x^*_5/2 &
x_{1,2}^*&-x_{1,2}^*&x_{4,1}^* &-x_{4,1}^*&x^*_3&x^*_3&x_{2,4}^*&-x_{2,4}^*\\
-x^*_5/2&-x^*_5/2&x^*_5/2&x^*_5/2&-x^*_5/2&-x^*_5/2&x^*_5/2&x^*_5/2&  
-x_{4,1}^*&-x_{4,1}^*&x_{1,2}^* &x_{1,2}^*&-x_{2,4}&-x_{2,4}&x_3&-x_3\\
-x^*_5/2& x^*_5/2& x^*_5/2 & -x^*_5/2 & -x^*_5/2 & x^*_5/2 &x^*_5/2 & -x^*_5/2 &
-x_{4,1}&x_{4,1}&x_{1,2} &-x_{1,2}&-x_{2,4}^*&x_{2,4}&x^*_3&x^*_3\\
-x^*_5/2&-x^*_5/2 & -x^*_5/2&-x^*_5/2 &x^*_5/2 &x^*_5/2 &x^*_5/2 &x^*_5/2   & 
-x_3&x_3&x_{2,4}^* &x_{2,4}^*&x_{1,2}&x_{1,2}&-x_{4,1}^*&x_{4,1}^*\\
 -x^*_5/2 & x^*_5/2 & -x^*_5/2 & x^*_5/2 & x^*_5/2 & -x^*_5/2 & x^*_5/2 & -x^*_5/2 & 
-x^*_3&-x^*_3&x_{2,4}& -x_{2,4}&x_{1,2}^*&-x_{1,2}^*&-x_{4,1}&x_{4,1}\\
-x^*_5/2&-x^*_5/2&x^*_5/2 & x^*_5/2 & x^*_5/2 & x^*_5/2 & -x^*_5/2 & -x^*_5/2 & 
-x_{2,4}^*&-x_{2,4}^*&-x_3 & x_3&x_{4,1}&x_{4,1}&x_{1,2}^*&x_{1,2}^*\\
-x^*_5/2 & x^*_5/2 & x^*_5/2 & -x^*_5/2 & x^*_5/2 & -x^*_5/2 & -x^*_5/2 & x^*_5/2 & 
-x_{2,4}&x_{2,4}&-x^*_3& -x^*_3&x_{4,1}^*&-x_{4,1}^*&x_{1,2}&-x_{1,2}\\
\end{array}
\right]
\end{equation}
\hrule
\end{figure*}
Observe that for a fixed average power per codeword, due to the presence of zeros in $G_3$, the peak power transmission in an antenna using $G_3$ will be higher than that of an antenna using $G_Y$. Hence, it is clear that the PAPR for the code $G_Y$ is lower than that of the code $G_3$. Hence, lower the fraction of zeros in a code, lower will be the PAPR of the code. In \cite{TWMS,STWWWXZ,ZSXWWWT}, another rate-$1/2,$ $8$ antenna code with no zero entry, denoted by $G_{TWMS}$ shown at the top of this page, has been reported. Observe that $G_{TWMS}$ has entries that are coordinated interleaved variables and hence has larger signaling complexity as explained in the following subsection.
\subsection{Signaling Complexity}
\label{subsec1_1}
The code given in ~\cite{TJC} obtained from Amicable Orthogonal Designs \cite{GeS}
\begin{equation}
\label{rcod4}
{\bf W}_{TJC}= \left[
\begin{array}{cccc}
s_1   & s_2   &\frac{s_3}{\sqrt{2}} & \frac{s_3}{\sqrt{2}}     \\
-s_2^* & s_1^* &\frac{s_3}{\sqrt{2}} & \frac{-s_3}{\sqrt{2}}   \\
\frac{s_3^*}{\sqrt{2}} & \frac{s_3^*}{\sqrt{2}}  &\frac{(-s_1-s_1^*+s_2-s_2^*)}{2}& \frac{(s_1-s_1^*-s_2-s_2^*)}{2} \\ \frac{s_3^*}{\sqrt{2}} & \frac{-s_3^*}{\sqrt{2}}  &\frac{(s_1-s_1^*+s_2+s_2^*)}{2} &-\frac{(s_1+s_1^*+s_2-s_2^*)}{2}
\end{array}
\right]
\end{equation}
is not a COD and the number of zeros in \eqref{rcod4} is zero. Notice that some of the entries of \eqref{rcod4} can be written as
\begin{equation}
\label{ciod}
\begin{array}{rcl}
\frac{(-s_1-s_1^*+s_2-s_2^*)}{2}=& -(s_{1I}-js_{2Q})=& -{s}_{1,2}^*; \\
\frac{(s_1-s_1^*-s_2-s_2^*)}{2} =& -(s_{2I}-js_{1Q})=& -{s}_{2,1}^*; \\
\frac{(s_1-s_1^*+s_2+s_2^*)}{2} =&s_{2I}+js_{1Q}    =&{s}_{2,1}; \\
-\frac{(s_1+s_1^*+s_2-s_2^*)}{2}=&-(s_{1I}+js_{2Q}) =& -{s}_{1,2}.
\end{array}
\end{equation}

The code ${\bf W}_{TJC}$ reported in \cite{TJC} is a NZE $4$-antenna code  and the NZE 4-antenna code ${\bf W}_{YGT}$ reported in \cite{YGT1} is
{\small
\begin{eqnarray}
\label{YGT4code}
\frac{1}{\sqrt{2}}\left[
\begin{array}{cccc}
s_1^*-s_2& s_1^*+s_2& s_3^*& -s_3^*\\ js_1+js_2^* & -js_1+js_2^*& js_3^*& js_3^*\\ -s_3 & s_3& s_1^*-s_2^*& s_1^*+s_2^*\\ -js_3 & -js_3& js_1+js_2& -js_1+js_2 \end{array}
\right].
\end{eqnarray}
}
It is important to note that whenever the code matrix has entries with more than one complex variable like 8 of the 16 entries in \eqref{YGT4code}, the number of possible transmitted values increases compared to having only one complex variable or its conjugate with or without negation. For example, if $s_1$ and $s_2$ take values from 16-QAM, 4 bits are needed to specify either one of them whereas 8 bits are needed to specify  $s_1^*-s_2.$ We say that the signaling complexity in specifying $s_1^*-s_2$ is more compared to specifying either $s_1$ or $s_2$ alone. In this sense, the signaling complexity of \eqref{YGT4code} is more than that of the code \eqref{itcod}.

Whenever coordinate interleaving appears, as in \eqref{rcod4}, some of the entries are of the form $s_{iI}\pm j s_{kQ}$ where $i \neq k.$ Now, suppose $s_1$ and $s_2$ take values from a unrotated square QAM constellation, say 16-QAM, $\{(x,y)| x,y \in \{ \pm1, \pm3  \} \}$  for concreteness and illustration purposes. To specify a value taken by $s_1,$ one needs two look-up tables with four entries each, one to specify $s_{1I}$ and the other to specify $s_{1Q}.$ To specify a coordinate interleaved term like $s_{1I}+js_{2Q}$ also one needs two look-up tables with four entries each, one to specify $s_{1I}$ and the other to specify $s_{2Q}.$ However, if one needs to rotate the 16-QAM constellation, for some purposes like guaranteeing full-diversity, then to specify the value taken by a term like $s_1$ one needs a look up table with 16 entries to specify $s_{1I}$ and  $s_{1I}$ uniquely specifies $s_{1Q}.$ This is true for coordinate interleaved terms also. Notice that a look up table with 16-entries need more memory/space than two look up tables with 4 entries each. In such cases also, we say that the signaling complexity increases.
Since  coordinate interleaving is a specific complex linear combination of variables as seen from \eqref{ciod} and designs using coordinate interleaving generally use rotated constellations for full-diversity and/or optimum coding gain, we say that designs that have entries that are linear combinations of several variables increase the signaling complexity of the design. Accordingly, the signaling complexity of the design given by \eqref{rcod4} is larger than that of the code for 4-antennas obtained from \eqref{itcod}. Notice that the signaling complexity of \eqref{YGT4code} is larger than that of \eqref{rcod4}, since there are 4 real variables involved in 8 entries of the matrix.

\subsection{Contributions}
Notice that by multiplying the matrix \eqref{itcod} with a unitary matrix the resulting matrix will continue to be a COD with lesser number of zeros and it is not difficult to locate unitary matrices that will result in a design with no zero entries. However, such a design is likely to have large signaling complexity which needs to be avoided. Obtaining  a unitary matrix which reduces the number of zero entries while not increasing the signaling complexity is a nontrivial task which has been attempted in \cite{DaR,DaR1} with partial success.
It is known that there always exist codes with no zero entry for $2^a$ transmit antennas if $a+1$ is a power of $2$ \cite{DaR}.
For example, for $8$ antennas, we have the scaled-COD $R_3$ with no zero entry

{\footnotesize 
\begin{equation*}
R_3=\frac{1}{\sqrt{2}}\left[\begin{array}{rrrrrrrr}
   x_1   &-x_2^*  &-x_3^*& x_4  &-x_4^* &-x_3   & x_2  & x_1^* \\
   x_1   &-x_2^*  &-x_3^*&-x_4  &-x_4^* & x_3   &- x_2 &- x_1^*\\
   x_2   & x_1^*  & x_4  &-x_3^*&-x_3   &-x_4^* & x_1  &-x_2^*\\
   x_2   & x_1^*  &-x_4  &-x_3^*& x_3   &-x_4^* &-x_1  & x_2^*\\
   x_3   & x_4    & x_1^*& x_2^*&-x_2   & x_1   &-x_4^*& x_3^*\\
   x_3   &-x_4    & x_1^*& x_2^*& x_2   &-x_1   &-x_4^*&-x_3^*\\
   x_4   & x_3    &-x_2  & x_1  & x_1^* & x_2^* & x_3^*&-x_4^* \\
  -x_4   & x_3    &-x_2  & x_1  &- x_1^*&- x_2^*&-x_3^*&-x_4^*
\end{array}\right].
\end{equation*}
}
\noindent
In general, for $2^a$ antennas, there exists a scaled-COD (denoted by $R_a$) with fraction of zeros equal to $(1-\frac{a+1}{2^a}2^{\lfloor log_2(\frac{2^a}{a+1}) \rfloor})$ for all $a$ \cite{DaR}. It is clear that the above quantity is not equal to zero if $a+1$ is not a power of $2$. It is therefore important to construct codes with no zero entry for $2^a$ antennas when $a+1$ is not a power of $2$. However, it is known there exists a code for $4$ transmit antennas with no zero entry \cite{DaR1} given by
\begin{eqnarray*}
\label{ldash2}
L_2=\left[\begin{array}{rrrr}
    x_1   &-x_2^*  &-\frac{x_3^*}{\sqrt{2}} &-\frac{x_3^*}{\sqrt{2}}\\
    x_2   & x_1^*  &-\frac{x_3^*}{\sqrt{2}} &\frac{x_3}{\sqrt{2}}\\
    \frac{x_3}{\sqrt{2}}   & \frac{x_3}{\sqrt{2}}    &x_{2,1}^*  &x_{1,2} \\
    \frac{x_3}{\sqrt{2}}   &-\frac{x_3}{\sqrt{2}}    & x_{1,2}^*  &-x_{2,1} 
\end{array}\right],
\end{eqnarray*}
Note that the signaling complexity of the above code is slightly more than that of the code $R_3$ as all the non-zero entries in $R_3$ are variables or its conjugates (upto scaling) while some of the entries in $L_2$ contains co-ordinate interleaved variables.
For $16$ transmit antennas, there also exist a code with no zero entry given by \eqref{NZE1} where $x_{i,k}=x_{iI}+jx_{kQ}$ \cite{PRD}. Other than $4$  and $16$ antennas, no code is known for $2^a$ antennas where $a+1$ is not a power of $2$. 
Note that, in $L_2,$ only $x_1$ and $x_2$ form co-ordinate interleaved variables denoted by $x_{1,2}, x_{2,1}$ whereas the other complex variable does not appear as coordinate interleaved with other variables. 
This particular observation is also valid for all the no zero entry designs constructed in this paper.  We will come to this observation later when the method for the construction of such codes is described.

In this paper, we provide a general procedure to construct SCODs with no zero entries for any power of two number of antennas, with marginal increase in the signaling complexity. Our contributions are summarized as follows:
\begin{itemize}
\item Maximal-rate square CODs with no zero entry for $2^a$ transmit antennas for any integer $a$.
\item Our construction is based on the multiplication of the code in \eqref{itcod} by a suitable pre-multiplying and a post-multiplying matrix consisting of only $\pm \frac{1}{\sqrt{\lambda}}$ or $0$ where $\lambda$ is a power of $2$ and hence easy to construct. We give a closed form expression for these pre- and post-multiplying matrices.

\item Only two variables of the design get coordinate interleaved and hence the increase in the signaling complexity compared to the one (if at all it existed) with no variables coordinate interleaved is very small. 
\end{itemize}

The remaining content of the paper is organized as follows: In Section \ref{sec2}, we prove the main result of the paper given by Theorem \ref{fracz} and discuss the signaling complexity of the constructed codes. Simulation results are given in Section \ref{sec3} and concluding remarks constitute Section \ref{sec4}.
\section{Construction of SCODs with no zero entry}
\label{sec2}
 In this section, we construct square CIS-CODs for any power of $2$ antennas such that all the entries in the matrix are non-zero. 

Our construction is based on the multiplication of the code given in \eqref{itcod} by a suitably chosen pre-multiplying and post-multiplying matrices so that the signaling complexity of the resulting code increases marginally when compared with the codes of \eqref{itcod}.
For illustration, there exists two unitary matrices $P,Q$ of order $16$ given by \eqref{pandq}, which when multiplied with $G_4$ give a code $PG_4Q$ given by \eqref{nze16} in which none of the entries is zero.
In both the matrices $P$ and $Q$, as well as in all the matrices throughout the paper, $-1$ is represented by simply the minus sign.

In order to construct codes for any power of $2$ number antennas with no zero entry, we introduce some notations:

Let $\mathbb{F}_2$ be the finite field with two elements denoted by $0$ and $1$ with addition denoted by $b_1\oplus b_2$ and multiplication denoted by $b_1b_2$ where $ b_1,b_2 \in \mathbb{F}_2$.

Let $B$ be a finite subset of the set of natural numbers with  $c$ being its largest element and $a$ being the smallest integer  such that $2^a>c$. We can always identify each element of $B$ with an element of $\mathbb{F}_2^a$ using the following correspondence:
$x\in B \leftrightarrow (x_{a-1},\cdots,x_0)\in\mathbb{F}_2^a$ such that $x=\sum_{j=0}^{a-1} x_j2^j,x_j\in \mathbb{F}_2$.
The all zero vector and all one vector in $\mathbb{F}_2^a$ are denoted by $\mathbf{0}$ and $\mathbf{1}$ respectively. For $x\in B$, $\norm{x}$ 
denotes the Hamming weight of $x$.
Let $x=(x_{a-1},\cdots,x_0),y=(y_{a-1},\cdots,y_0),x_i,y_i\in \f_2$ for $i=0,1,\cdots,a-1$. Let 
$x\oplus y$ denote the component-wise modulo-2 addition of $x$ and $y$ respectively i.e.,
\begin{eqnarray*}
x\oplus y=(x_{a-1}\oplus y_{a-1},\cdots,x_0\oplus y_0).
\end{eqnarray*}
Let $Z_l=\{0,1,\cdots,l-1\}.$ We identify $Z_{2^a}$ with the set of $a$-tuple binary vectors $\f_2^a$ in the standard way, i.e., 
any element of $Z_{2^a}$ is identified with its radix-2 representation vectors (of length $a$). 
For convenience, the  set $Z_{2^a}$ is used as a collection of positive integers and sometimes as the set of vectors. 
For a set $K\subset Z_{2^a}$ and $m\in Z_{2^a}$, let $\vert K\vert$ be the number of elements in the set $K$ and $m \oplus K:=\{m\oplus a~\vert~ a\in K\}$.
For two sets $A$ and $B$, let $A\setminus B=\{x\in A\vert x\notin B\}.$ 
For two matrices $A=[a_{ij}]$ and $B$, the tensor product of $A$ with $B$, denoted by $A\otimes B$, is the matrix
$[a_{i,j}B]$.
For $\alpha,\beta$ positive integers with $\beta>\alpha$, define $[\alpha,\beta]:=\{\alpha,\alpha+1,\cdots,\beta\}$.

In the following, we construct the no zero entry code for $2^a$ antennas in two steps: First,
(i) we construct a code $K_a$ from $G_a$ such that the number of non-zero entries in $K_a$ is a power of $2$ and, then, 
(ii) we construct a code $L_a$ with no zero entry from $K_a$.\\

Define $b=\lfloor log_2({a})\rfloor+1,$ $m=2^b-a-1$ and $q=a-2^{b-1}$.
It is clear that for all $x\leq a$, we can express $x$ as $x=\sum_{j=0}^{b-1}x_j2^j$ with $x_j\in \mathbb{F}_2$.
Let
{\small
\begin{eqnarray*}
P_a&=&\{0,2^0,2^1,\cdots,2^{a-1}\},\\
Q_a &=&
    \begin{cases}
         \phi   \hspace{0.2 in}  \text{ if $a+1$ is a power of $2$, }&  \\
\{1\oplus 2^{a-m},1\oplus 2^{a-m+1},\cdots, 1\oplus2^{a-1}\}   & \text{ otherwise, } 
    \end{cases} 
\end{eqnarray*}
}
and 
\begin{eqnarray}
\label{Tdasha}
T_a=P_a\cup Q_a, \nonumber\\
T_a^{(i)}=i\oplus T_a \mbox{ for all }i\in Z_{2^a}.
\end{eqnarray}

Note that $\abs{T_a^{(i)}}=2^b$ for all $i$.
\begin{figure*}
 {\small
\begin{equation}
\label{pandq}
P=\frac{1}{\sqrt{2}}
\left[
\begin{array}{ r @{\hspace{.2pt}} r @{\hspace{.2pt}} r @{\hspace{.2pt}} r @{\hspace{.2pt}} r @{\hspace{.2pt}} r @{\hspace{.2pt}} r @{\hspace{.2pt}} r @{\hspace{.2pt}} r @{\hspace{.2pt}} r @{\hspace{.2pt}} r @{\hspace{.2pt}} r @{\hspace{.2pt}} r @{\hspace{.2pt}} r @{\hspace{.2pt}} r @{\hspace{.2pt}} r @{\hspace{.2pt}} r @{\hspace{.2pt}} r @{\hspace{.2pt}} r @{\hspace{.2pt}} r @{\hspace{.2pt}} r @{\hspace{.2pt}} r @{\hspace{.2pt}} r @{\hspace{.2pt}} r  @{\hspace{.2pt}}r @{\hspace{.2pt}} r @{\hspace{.2pt}} r @{\hspace{.2pt}} r @{\hspace{.2pt}} r @{\hspace{.2pt}} r @{\hspace{.2pt}} r @{\hspace{.2pt}} r } 
  \sqrt{2}&  ~0&  ~0&  ~0&  ~0&  ~0&  ~0&  ~0&  ~0&  ~0&  ~0&  ~0&  ~0&  ~0&  ~0&  ~0 \\
  0&  \sqrt{2}&  0&  0&  0&  0&  0&  0&  0&  0&  0&  0&  0&  0&  0&  0\\
  0&  0&  1&  1&  0&  0&  0&  0&  0&  0&  0&  0&  0&  0&  0&  0\\
  0&  0&  1& -&  0&  0&  0&  0&  0&  0&  0&  0&  0&  0&  0&  0\\
  0&  0&  0&  0&  1&  1&  0&  0&  0&  0&  0&  0&  0&  0&  0&  0\\
  0&  0&  0&  0&  1& -&  0&  0&  0&  0&  0&  0&  0&  0&  0&  0\\
  0&  0&  0&  0&  0&  0&  \sqrt{2}&  0&  0&  0&  0&  0&  0&  0&  0&  0\\
  0&  0&  0&  0&  0&  0&  0&  \sqrt{2}&  0&  0&  0&  0&  0&  0&  0&  0\\
  0&  0&  0&  0&  0&  0&  0&  0&  1&  1&  0&  0&  0&  0&  0&  0\\
  0&  0&  0&  0&  0&  0&  0&  0&  1& -&  0&  0&  0&  0&  0&  0\\
  0&  0&  0&  0&  0&  0&  0&  0&  0&  0&  \sqrt{2}&  0&  0&  0&  0&  0\\
  0&  0&  0&  0&  0&  0&  0&  0&  0&  0&  0&  \sqrt{2}&  0&  0&  0&  0\\
  0&  0&  0&  0&  0&  0&  0&  0&  0&  0&  0&  0&  \sqrt{2}&  0&  0&  0\\
  0&  0&  0&  0&  0&  0&  0&  0&  0&  0&  0&  0&  0&  \sqrt{2}&  0&  0\\
  0&  0&  0&  0&  0&  0&  0&  0&  0&  0&  0&  0&  0&  0&  1&  1\\
  0&  0&  0&  0&  0&  0&  0&  0&  0&  0&  0&  0&  0&  0&  1& -
\end{array}
\right],~~
Q=\frac{1}{2}
\left[ \begin{array}{ r @{\hspace{.2pt}} r @{\hspace{.2pt}} r @{\hspace{.2pt}} r @{\hspace{.2pt}} r @{\hspace{.2pt}} r @{\hspace{.2pt}} r @{\hspace{.2pt}} r @{\hspace{.2pt}} r @{\hspace{.2pt}} r @{\hspace{.2pt}} r @{\hspace{.2pt}} r @{\hspace{.2pt}} r @{\hspace{.2pt}} r @{\hspace{.2pt}} r @{\hspace{.2pt}} r @{\hspace{.2pt}} r @{\hspace{.2pt}} r @{\hspace{.2pt}} r @{\hspace{.2pt}} r @{\hspace{.2pt}} r @{\hspace{.2pt}} r @{\hspace{.2pt}} r @{\hspace{.2pt}} r  @{\hspace{.2pt}}r @{\hspace{.2pt}} r @{\hspace{.2pt}} r @{\hspace{.2pt}} r @{\hspace{.2pt}} r @{\hspace{.2pt}} r @{\hspace{.2pt}} r @{\hspace{.2pt}} r } 
 \sqrt{2}&  ~0&  ~0&  ~0&  ~0&  ~0&  ~0&  ~0&  ~0&  ~0&  ~0&  ~0&  ~0&  ~0&  ~1&~-\\
 \sqrt{2}&  0&  0&  0&  0&  0&  0&  0&  0&  0&  0&  0&  0&  0&-&  1\\
  0& \sqrt{2}&  0&  0&  0&  0&  0&  0&  0&  0&  0&  0&  0&  0&  1&  1\\
  0& \sqrt{2}&  0&  0&  0&  0&  0&  0&  0&  0&  0&  0&  0&  0&-&-\\
  0&  0&  1&  1&  0&  0&  0&  0&  0&  0&  0&  0&  0& \sqrt{2}&  0&  0\\
  0&  0&  1&  1&  0&  0&  0&  0&  0&  0&  0&  0&  0&-\sqrt{2}&  0&  0\\
  0&  0&  1&-&  0&  0&  0&  0&  0&  0&  0&  0& \sqrt{2}&  0&  0&  0\\
  0&  0&  1&-&  0&  0&  0&  0&  0&  0&  0&  0&-\sqrt{2}&  0&  0&  0\\
  0&  0&  0&  0&  1&  1&  0&  0&  0&  0&  0& \sqrt{2}&  0&  0&  0&  0\\
  0&  0&  0&  0&  1&  1&  0&  0&  0&  0&  0&-\sqrt{2}&  0&  0&  0&  0\\
  0&  0&  0&  0&  1&-&  0&  0&  0&  0& \sqrt{2}&  0&  0&  0&  0&  0\\
  0&  0&  0&  0&  1&-&  0&  0&  0&  0&-\sqrt{2}&  0&  0&  0&  0&  0\\
  0&  0&  0&  0&  0&  0& \sqrt{2}&  0&  1&-&  0&  0&  0&  0&  0&  0\\
  0&  0&  0&  0&  0&  0& \sqrt{2}&  0&-&  1&  0&  0&  0&  0&  0&  0\\
  0&  0&  0&  0&  0&  0&  0& \sqrt{2}&  1&  1&  0&  0&  0&  0&  0&  0\\
  0&  0&  0&  0&  0&  0&  0& \sqrt{2}&-&-&  0&  0&  0&  0&  0&  0
\end{array}
\right]
\end{equation}
}
\hrule
\end{figure*}
\begin{figure*}
{ \small
\begin{equation}
\hspace{-10pt}
\label{nze16}
\frac{1}{2}\left[ \hspace{-5pt}
\begin{array}{ r @{\hspace{.2pt}} r @{\hspace{.2pt}} r @{\hspace{.2pt}} r @{\hspace{.2pt}} r @{\hspace{.2pt}} r @{\hspace{.2pt}} r @{\hspace{.2pt}} r @{\hspace{.2pt}} r @{\hspace{.2pt}} r @{\hspace{.2pt}} r @{\hspace{.2pt}} r @{\hspace{.2pt}} r @{\hspace{.2pt}} r @{\hspace{.2pt}} r @{\hspace{.2pt}} r @{\hspace{.2pt}} r @{\hspace{.2pt}} r @{\hspace{.2pt}} r @{\hspace{.2pt}} r @{\hspace{.2pt}} r @{\hspace{.2pt}} r @{\hspace{.2pt}} r @{\hspace{.2pt}} r  @{\hspace{.2pt}}r @{\hspace{.2pt}} r @{\hspace{.2pt}} r @{\hspace{.2pt}} r @{\hspace{.2pt}} r @{\hspace{.2pt}} r @{\hspace{.2pt}} r @{\hspace{.2pt}} r } 
 \sqrt{2}x_1&     -\sqrt{2}x_2^*&   -x_3^*&   -x_3^*&   -x_4^*&   -x_4^*& x_5&-x_5&   -x_5^*&   -x_5^*&-x_4& x_4& x_3&-x_3&  \sqrt{2}{x}_{2,1}^*&  \sqrt{2}{x}_{1,2}\\
 \sqrt{2}x_1&     -\sqrt{2}x_2^*&   -x_3^*&   -x_3^*&   -x_4^*&   -x_4^*&-x_5& x_5&   -x_5^*&   -x_5^*& x_4&-x_4&-x_3& x_3& -\sqrt{2}{x}_{2,1}^*& -\sqrt{2}{x}_{1,2}\\
 \sqrt{2}x_2&      \sqrt{2}x_1^*&   -x_3^*&    x_3^*&   -x_4^*&    x_4^*& x_5& x_5&   -x_5^*&    x_5^*&-x_4&-x_4& x_3& x_3&\sqrt{2}{x}_{1,2}^*&-\sqrt{2}{x}_{2,1}\\
 \sqrt{2}x_2&      \sqrt{2}x_1^*&   -x_3^*&    x_3^*&   -x_4^*&    x_4^*&-x_5&-x_5&   -x_5^*&    x_5^*& x_4& x_4&-x_3&-x_3& -\sqrt{2}{x}_{1,2}^*& \sqrt{2}{x}_{2,1}\\
x_3& x_3&\sqrt{2}{x}_{1,2}^*&-\sqrt{2}{x}_{2,1}& x_5&-x_5&   -x_4^*&   -x_4^*&-x_4& x_4&   -x_5^*&   -x_5^*&  \sqrt{2}x_2&      \sqrt{2}x_1^*&   -x_3^*&    x_3^*\\
x_3& x_3&\sqrt{2}{x}_{1,2}^*&-\sqrt{2}{x}_{2,1}&-x_5& x_5&   -x_4^*&   -x_4^*& x_4&-x_4&   -x_5^*&   -x_5^*& -\sqrt{2}x_2&     -\sqrt{2}x_1^*&    x_3^*&   -x_3^*\\
x_3&-x_3&  \sqrt{2}{x}_{2,1}^*&  \sqrt{2}{x}_{1,2}& x_5& x_5&   -x_4^*&    x_4^*&-x_4&-x_4&   -x_5^*&    x_5^*&  \sqrt{2}x_1&     -\sqrt{2}x_2^*&   -x_3^*&   -x_3^*\\
x_3&-x_3&  \sqrt{2}{x}_{2,1}^*&  \sqrt{2}{x}_{1,2}&-x_5&-x_5&   -x_4^*&    x_4^*& x_4& x_4&   -x_5^*&    x_5^*& -\sqrt{2}x_1&      \sqrt{2}x_2^*&    x_3^*&    x_3^*\\
x_4& x_4& x_5&-x_5&\sqrt{2}{x}_{1,2}^*&-\sqrt{2}{x}_{2,1}&    x_3^*&    x_3^*&-x_3& x_3&  \sqrt{2}x_2&      \sqrt{2}x_1^*&   -x_5^*&   -x_5^*&    x_4^*&   -x_4^*\\
x_4& x_4&-x_5& x_5&\sqrt{2}{x}_{1,2}^*&-\sqrt{2}{x}_{2,1}&    x_3^*&    x_3^*& x_3&-x_3& -\sqrt{2}x_2&     -\sqrt{2}x_1^*&   -x_5^*&   -x_5^*&   -x_4^*&    x_4^*\\
x_4&-x_4& x_5& x_5&  \sqrt{2}{x}_{2,1}^*&  \sqrt{2}{x}_{1,2}&    x_3^*&   -x_3^*&-x_3&-x_3&  \sqrt{2}x_1&     -\sqrt{2}x_2^*&   -x_5^*&    x_5^*&    x_4^*&    x_4^*\\
x_4&-x_4&-x_5&-x_5&  \sqrt{2}{x}_{2,1}^*&  \sqrt{2}{x}_{1,2}&    x_3^*&   -x_3^*& x_3& x_3& -\sqrt{2}x_1&      \sqrt{2}x_2^*&   -x_5^*&    x_5^*&   -x_4^*&   -x_4^*\\
x_5&-x_5& x_4& x_4&-x_3&-x_3&  \sqrt{2}x_1&     -\sqrt{2}x_2^*&  \sqrt{2}{x}_{2,1}^*&  \sqrt{2}{x}_{1,2}&    x_3^*&   -x_3^*&    x_4^*&   -x_4^*&   -x_5^*&   -x_5^*\\
-x_5& x_5& x_4& x_4&-x_3&-x_3&  \sqrt{2}x_1&     -\sqrt{2}x_2^*& -\sqrt{2}{x}_{2,1}^*& -\sqrt{2}{x}_{1,2}&   -x_3^*&    x_3^*&   -x_4^*&    x_4^*&   -x_5^*&   -x_5^*\\
x_5& x_5& x_4&-x_4&-x_3& x_3&  \sqrt{2}x_2&      \sqrt{2}x_1^*&\sqrt{2}{x}_{1,2}^*&-\sqrt{2}{x}_{2,1}&    x_3^*&    x_3^*&    x_4^*&    x_4^*&   -x_5^*&    x_5^*\\
-x_5&-x_5& x_4&-x_4&-x_3& x_3&  \sqrt{2}x_2&      \sqrt{2}x_1^*&-\sqrt{2}{x}_{1,2}^*& \sqrt{2}{x}_{2,1}&   -x_3^*&   -x_3^*&   -x_4^*&   -x_4^*&   -x_5^*&    x_5^*\\
\end{array}
\right]
\end{equation}
}
\hrule
\end{figure*}
With  
 \begin{eqnarray*}
\begin{array}{rlc}
W_a &=& \left[\begin{array}{cccc}
  A_0      &0         & \cdots &0         \\
  0        &A_1      & \cdots &0         \\
 \vdots    &\vdots    & \ddots &\vdots    \\
  0        &0         & \cdots &A_{2^m-1} 
\end{array}\right],\\
\end{array}
\end{eqnarray*}
\begin{equation}
\label{dkdasha}
K_a=W_aG_aW_a,
\end{equation}
where 
  \begin{eqnarray*}
\begin{array}{c}
    A_i =
    \begin{cases}
         I_{2^{a-m}}          & \text{ if $\norm{i}$ is even , } \\
         I_{2^{a-m-1}}\otimes (\frac{1}{\sqrt{2}}H_2)          & \text{ if $\norm{i}$ is odd } 
    \end{cases} 
\end{array}
\end{eqnarray*}
and $H_2= \left[\begin{array}{rr} 1 & 1 \\ -1 & 1    \end{array} \right].$

One nice property of the matrix $K_a$ is that the number of non-zero entries in $K_a$ is a power of $2$. Let $N_i^{(G_a)},N_i^{(K_a)}$ be the set of the column indices of the non-zero entries in the $i$-th row of $G_a$ and $K_a$ respectively.
It is known \cite{DaR} that 
$N_i^{(G_a)}=\{i\}\cup\{i \oplus 2^j~\vert~ j=0 \mbox { to } a-1\}$.

 The following lemma describes the set $N_i^{(K_a)}$.
\begin{table*}
\caption{$M_a,\tilde{M}_a$ and $M_a^\prime$ for $a=3,4,\cdots,9$ } 
\begin{center}
\begin{tabular}{|c|c|c|c|c|c|c|c|} \hline \label{tab1}
$a$ & 3 & 4& 5& 6 & 7 & 8  & 9  \\ \hline
$M_a$  &$\{3\}$ &$\{3\}$&$\{3,5\}$&$\{3,5,6\}$ & $\{3,5,6,7\}$ & $\{3,5,6,7\}$& $\{3,5,6,7,9\}$  \\ \hline
$\tilde{M}_a$ &$\{3\}$ &$\{6\}$ &$\{3,6\}$&$\{3,5,6\}$ & $\{3,5,6,7\}$ & $\{6,10,12,14\}$ & $\{3,6,10,12,14\}$  \\ \hline
$M_a^\prime$ & $\{7\}$ & $\{14\}$ & $\{7,26\}$ & $\{7,25,42,\}$ & $\{7,25,42,75\}$ &$\{42,134,152,202\}$ &$\{7,42,134,152,202\}$ \\ \hline
$b$ &2 & 3&3&3&3 &4&4\\ \hline
\end{tabular}
\end{center}
\end{table*}
\begin{lemma}
\label{kdasha}
Let $a$ be a positive integer, $s\in Z_{2^a}$ and $T_a^{(s)}$ be as given by \eqref{Tdasha}. Then
$N_s^{(K_a)}=T_a^{(s)}$.
\end{lemma}

\begin{proof}
Let 
\begin{equation*}
K_a = \left[\begin{array}{cccc}
 K_{0,0}      &K_{0,1}         & \cdots &K_{0,2^m-1}         \\
K_{1,0}      &K_{1,1}         & \cdots &K_{1,2^m-1}         \\
\vdots    &\vdots    & \ddots &\vdots    \\
K_{2^{m}-1,0}      &K_{2^m-1,1}         & \cdots &K_{2^m-1,2^m-1}         \\
\end{array}\right]
\end{equation*}
\noindent
where $K_{i,j}$ is a square matrix of order $2^{a-m}$ for $0\leq i,j\leq 2^m-1$.
Similarly, we write $G_a$ in the above form  and let $(i,j)$-th block matrix of $G_a$
be $G_{i,j}$.
We have $K_{i,j}=A_iG_{i,j}A_j$ and  

{\footnotesize
  \begin{eqnarray*}
\begin{array}{c}
    K_{i,j} =
    \begin{cases}
       G_{i,j}  & \text{ if $\norm{i}$ even, $\norm{j}$ even } \\
       G_{i,j}(I_{2^{a-m-1}}\otimes H_2)       & \text{ if $\norm{i}$ even, $\norm{j}$ odd } \\
       (I_{2^{a-m-1}}\otimes H_2)G_{i,j}       & \text{ if $\norm{i}$ odd, $\norm{j}$ even } \\
      (I_{2^{a-m-1}}\otimes H_2)G_{i,j}(I_{2^{a-m-1}}\otimes H_2)       & \text{ if $\norm{i}$ odd, $\norm{j}$ odd } 
    \end{cases}. 
\end{array}
\end{eqnarray*}
}

We now compute $N_s^{(K_a)}$ as follows:
Let $s=2^{a-m}r+t$. 
We now consider two cases: (i) $\norm{r}$ even and (ii) $\norm{r}$ odd. \\
Let $\alpha=2^{a-m}j$ and $\beta=2^{a-m}j+2^{a-m}-1$.\\
For the first case, we have
  \begin{eqnarray*}
 N_s^{(K_a)}&=&
\left( \bigcup_{\substack{j=0\\ \norm{j} even}} ^{2^m-1 }
(N_s^{(G_a)}\cap [\alpha,\beta]) \right) \bigcup \\
&& \left(\bigcup_{\substack{j=0\\ \norm{j} odd}} ^{2^m-1 }
 \left((N_s^{(G_a)}\cup N_{s\oplus 1}^{(G_a)})\cap [\alpha,\beta]\right) \right)\\
  &=& N_s^{(G_a)}\cup Z
\end{eqnarray*}
where 
$$Z=\bigcup_{\substack{j=0\\ \norm{j} odd}} ^{2^m-1 } 
\left (N_{s\oplus 1}^{(G_a)}\cap [\alpha,\beta]\right)  $$
By simplifying the above expression, we have
 $Z= s\oplus 1\oplus \{2^{a-m},2^{a-m+1},\cdots,2^{a-1}\}$.
Hence $N_s^{(K_a)}=T_a^{(s)}$.
The proof for the case when $\norm{s}$ is odd is similar.
\end{proof}
Two distinct rows of $K_a$, say the $s$'th and the $t$'th are said to be 
{\it{non-intersecting}} if $N_s^{(K_a)}\cap N_t^{(K_a)}=\phi$.

Let $M_a=\{0<x\leq a~\vert ~ x\neq 2^k \text { for any $k=0,1,\cdots$ }\}$. For all $x\in M_a$, write $x=\sum_{j=0}^{b-1}x_j2^j$.
Define a function $g$ on $M_a$ as follows:
\begin{eqnarray*}
\begin{array}{c}
      g(x) =
    \begin{cases}
          2x          & \text{ if $x_{b-1}=0$, }\\
          2x+1-2^b  & \text{ if $x_{b-1}=1$ }.
    \end{cases}
\end{array}
\end{eqnarray*}
\begin{lemma}
\label{mtilde}
Let $x\in M_a$ be such that $x_{b-1}=1$. Then $g(x)-1< a-m$.
\end{lemma}
\begin{proof}
As $x_{b-1}=1$, we have $g(x)=2x+1-2^b$.
Therefore, 
\begin{eqnarray*}
a-m-g(x)+1&=&a-(2^b-a-1)-(2x+1-2^b)+1\\
         &&=2a+1-2x\geq 1>0
\end{eqnarray*}
as $x\leq a$.
\end{proof}
Let
\begin{eqnarray*}
\label{madash}
\tilde{M}_a=\{g(x)~\vert ~ x\in M_a\}.
\end{eqnarray*}
Note that $\tilde{M}_a=M_a$ for $a=2^{b}-2,2^{b}-1$. For all other values of $a$, $\tilde{M}_a\neq M_a$. 
Let $J_a= M_a\setminus \tilde{M}_a$ and $H_a=\tilde{M}_a\setminus M_a$.
It is clear that $J_a=H_a=\phi$ for $a=2^b-2,2^{b}-1$. For all other values of $a$, we have 
\begin{eqnarray}
\label{ma1}
\begin{array}{lcl}
H_a&=&\{2\lceil \frac{a+1}{2}\rceil, 2(\lceil \frac{a+1}{2}\rceil+1),\cdots,
2(2^{b-1}-1)\},\\
J_a&=&\{2q+3,2q+5,\cdots, 2\lceil \frac{a}{2} \rceil-1\}.
\end{array}
\end{eqnarray}
Note that $J_a\vert=\vert H_a\vert=\lceil \frac{m-1}{2} \rceil$ if $J_a\neq \phi$ and hence $H_a\neq \phi$.
Define $f^\prime:\tilde{M}_a\rightarrow M_a$ as follows:
\begin{eqnarray*}
\begin{array}{c}
      f^\prime(x) =
    \begin{cases}
          x                               & \text{ if $x\in M_a\cap\tilde{M}_a$ }\\
          x+1-2\lceil \frac{m}{2}\rceil & \text{ if $x\in H_a$ }.
    \end{cases}
\end{array}
\end{eqnarray*}
Note the map $f^\prime$ is well defined as $f^\prime(\tilde{M}_a)=M_a$ which follows from the fact that  $f^\prime(H_a)=J_a$. Moreover, the map $f^\prime$ is injective and $f^\prime\leq x$ for all $x\in \tilde{M}_a$.

Let $f:M_a\rightarrow M_a$ given by $f=f^\prime g$. It is a bijective map from $M_a$ to itself. 
Note that (i) $f(x)\leq g(x)$ for all $x\in M_a$ as $f^\prime\leq x$ for all $x\in \tilde{M}_a$ and (ii) $f(x)=g(x)$ if $x_{b-1}=1$ because $g(x)$ is odd if $x_{b-1}=1$ and hence $g(x)\notin H_a$. 

Now we define another function $h$ on $M_a$ as
\begin{eqnarray}
h(x)=2^{f(x)-1}+\sum_{j=0}^{b-2}x_j2^{2^{j+1}-1}+x_{b-1}
\end{eqnarray}
for all $x\in M_a$. The map $h$ is injective.
Let $$M_a^\prime=\Big\{ h(x) ~\Big\vert ~ x \in M_a\Big\}.$$
Consider $M_a^\prime$ as a subset of $\f_2^a$. 
Note that the number of elements in $M_a$ and also of $M_a^\prime$ is $a-b$. Let $S$ be the linear subspace of $\mathbb{F}_2^a$ spanned by the elements of $M_a^\prime$. As the elements of $M_a^\prime$ are linearly independent over $\mathbb{F}_2$, the dimension of $S$ is $a-b$.
\begin{example}
The sets $M_a,\tilde{M}_a$ and $M_a^\prime$ for $a=3,4,\cdots,9$ are shown in Table \ref{tab1} at the top of this page.
\end{example}
Let 
\begin{eqnarray*}
2_+^x&=&
    \begin{cases}
       2^x    & \mbox{ if } x \geq 0,\\
       0      & \mbox{ if } x=-1. 
    \end{cases}
\end{eqnarray*}
(The situation where $x < -1$  never arise throughout the paper.)
\begin{lemma}
\label{pdist3}
Let $S$ be as above. Then $T_a^{(i)}\cap T_a^{(j)}=\phi$ for all $i,j\in S,i\neq j$.
\end{lemma}
\begin{proof}
Let $U_a=\{x\oplus y ~\vert~ x,y\in T_a\}$.
Any element in $U_a$ is one of the following types:\\
{\bf Type-I:} $2_+^\alpha+2_+^{\alpha^\prime},\alpha< \alpha^\prime$, (hence, $\alpha^\prime\neq -1$), \\
{\bf Type-II:} $1\oplus 2_+^\alpha+2_+^{\alpha^\prime},\alpha< \alpha^\prime,~\alpha\neq -1$, ( hence $\alpha^\prime\neq -1)$ such that $\alpha^\prime\geq a-m$.\\
One can easily check that $T_a^{(i)}\cap T_a^{(j)}=\phi$ for all $i,j\in S,i\neq j$ if and only if $x\notin U_a$ for all $x\in S$.
Therefore, it is enough to prove that 
(i)  the minimum Hamming distance (MHD) of $S$ is $3$ and 
(ii) if an element of $S$ is 
$1\oplus 2_+^\alpha+2_+^{\alpha^\prime},\alpha< \alpha^\prime,~\alpha\neq -1$, then $\alpha^\prime<a-m$.\\
{\bf Proof for Type-I:}  Note that $S=\{\sum_{j=0}^{a-b-1}c_jy_j^\prime ~\vert ~ y_j^\prime\in M_a^\prime \}$ where $c_j\in\f_2$ for $j=0,1,\cdots,{a-b-1}$, 
and the map $h$ given by
\begin{eqnarray}
\label{xdash}
\begin{array}{l}
 \hspace{40pt}      h : M_a \rightarrow M_a^\prime   \\
             x=\sum_{j=0}^{b-1} x_j2^j  \\ 
\mapsto  x^\prime= 2^{f(x)-1}+\sum_{j=0}^{b-2}x_j2^{2^{j+1}-1}+x_{b-1}. 
\end{array}
\end{eqnarray}
is one-one.
Now $2^{f(x)-1}\neq 2^{2^j-1}$ for $j=0,1,\cdots,b-1$ as $f(x)\neq 2^j$ for all $x\in M_a$.
Therefore, $\norm{x^\prime}=1+\norm{x}$ for all $x\in M_a$ where $\norm{x}$ stands for the Hamming weight of $x$.
But $\norm{x}\geq 2$ as $x$ is not a power of $2$, hence $\norm{x^\prime}\geq 3$.

Similarly, $\norm{x^\prime\oplus y^\prime}=2+ \norm{x\oplus y}$ for all $x,y\in M_a, x\neq y$.
Now $\norm{x\oplus y}\geq 1$ as $x\neq y$, which implies that $\norm{x^\prime\oplus y^\prime }\geq 3$ for all $x^\prime,y^\prime\in M_a^\prime$.
In general, $\norm{y_1^\prime \oplus y_2^\prime\oplus\cdots \oplus y_k^\prime}= k+ \norm{y_1 \oplus y_2\oplus \cdots \oplus y_k}$ for $k\leq a-d$, $y_1^\prime \neq y_2^\prime\neq \cdots \neq y_k^\prime$.
So for all $k\geq 3$ and $k\leq a-d$, $\norm{y_1^\prime \oplus y_2^\prime\oplus\cdots \oplus y_k^\prime}\geq 3$. As $\norm{h(x)}=3$ for $x=3$, the MHD of $S$ is $3$.\\
{\bf Proof for Type-II :}\\
Let $y=\sum_{j=0}^{a-b-1}c_jy_j^\prime,c_j\in \f_2,y_j^\prime\in
 M_a^\prime,j=0,1,\cdots,a-b-1$. If the Hamming weight of $y$ is $3$, then number of
 non-zero coefficients in the expansion of $y$ with respect to the elements of
 $M_a^\prime$ is atmost $3$.
We consider the following three cases:\\
(i) $y=y_1$,\\
(ii) $y=y_1\oplus y_2$ \\
(iii) $y=y_1\oplus y_2\oplus y_3$ where $y_i\in M_a^\prime,i=1,2,3$.\\

{\it Case (i):} \\
Let $y=1\oplus 2^{\alpha}\oplus 2^{\alpha^\prime}\in M_a^\prime$ with $\alpha^\prime>\alpha$. Then $y=h(z)$ for some $z\in M_a$ such that $z_{b-1}=1$ and $\norm{z}=2$.
Therefore, $z=2^{b-1}+2^\beta$ for some $\beta\in\{0,1,\cdots,b-2\}$. Moreover, $z\leq a$ and hence $2^\beta\leq a-2^{b-1}$.

Now $y=h(z)=2^{f(2^{b-1}+2^\beta)-1}+2^{2^{\beta+1}-1}+1$.
Now $max\{f(2^{b-1}+2^\beta),2^{\beta+1}\}=f(2^{b-1}+2^\beta)$ as $f(2^{b-1}+2^\beta)=2^{\beta+1}+1$. Therefore, $\alpha^\prime=f(2^{b-1}+2^\beta)-1$.
But $f(2^{b-1}+2^\beta)=g(2^{b-1}+2^\beta)$ and $g(x)-1<a-m$ by Lemma \ref{mtilde}.
Therefore, $\alpha^\prime<a-m$.\\

{\it Case (ii): }\\
Let $y=r^\prime+s^\prime$ for some $r^\prime,s^\prime\in M_a^\prime$ where $r^\prime=h(r),s^\prime=h(s)$, $r,s\in M_a$. 
Let $r=\sum_{j=0}^{b-1} r_j2^j,s=\sum_{j=0}^{b-1} s_j2^j$.
We have $y=2^{f(r)-1}\oplus\sum_{j=0}^{b-2} r_j2^{2^{j+1}-1}\oplus r_{b-1}\oplus 2^{f(s)-1}\oplus\sum_{j=0}^{b-2} s_j2^{2^{j+1}-1}\oplus s_{b-1}$.
Now $y=1\oplus 2^{\alpha}\oplus 2^{\alpha^\prime}$ which implies that 
$y=2^{f(r)-1}\oplus 2^{f(s)-1}\oplus 1$ as $f$ is injective and
$f(t)$ is not a power of $2$ for all $t\in M_a$.
Therefore, $r_j=s_j$ for $j=0,1,\cdots,b-2$ and $r_{b-1}\oplus s_{b-1}=1$.

Without loss of generality, we can assume that $r_{b-1}=0,s_{b-1}=1$, therefore $s>r$ and $g(s)$ is odd while $g(r)$ is even. Moreover, $g(s)=g(r)+1$.
Now $f(s)=g(s)$ if $s_{b-1}=1$ and $f(r)\leq g(r)$.
Therefore $max\{f(r),f(s)\}=f(s)$.
Now by Lemma \ref{mtilde}, $f(s)-1=g(s)-1<a-m$.\\

{\it Case (iii):}\\
Suppose $y=y_1^\prime\oplus y_2^\prime\oplus y_3^\prime$ where $y_i^\prime=h(y_i)$ for some $y_i\in M_a, i=1,2,3$.
As the Hamming weight of $y$ is $3$, we must have $y=2^{f(y_1)-1}\oplus 2^{f(y_2)-1}\oplus 2^{f(y_3)-1}$. If $y$ is of the form $y=1\oplus 2^{\alpha}\oplus 2^{\alpha^\prime}$, then  $f(y_i)=1$ for some $i\in\{1,2,3\}$ which is not true as $1\notin M_a$.
\end{proof}

\begin{lemma}
\label{part}
Let $a$ be a non-zero positive integer and $b$ be a positive integer such that $2^{b-1}\leq a<2^b$.
Then, there exists a partition of  $Z_{2^a}$ into $2^b$ subsets $C_j^{(a)}, j=0,1,\cdots,2^b-1$ each containing $2^{a-b}$ elements, such that 
$T_a^{(x)}\cap T_a^{(y)}=\phi$
for any two distinct elements $x,y\in C_j^{(a)}$ for any $j\in\{0,1,\cdots,2^b-1\}$.
\end{lemma}

\begin{proof}
We identify the set $Z_{2^a}$ with $\f_2^a$ as before.
Let $M_a^\prime$ be as given by \eqref{madash}
and  $S$ be the sub-space of $Z_{2^a}$ spanned by 
the elements of $M_a^\prime$. 
We define a relation $'\sim'$ on $Z_{2^a}$ as follows:
For all $\alpha,\beta\in Z_{2^a}$, $\alpha\sim \beta$, if $\alpha\oplus \beta\in S$.
One can easily check that this relation is an equivalence relation.
Moreover, the number of elements in any equivalence class is $2^{a-b}$ and hence 
the number of equivalence classes  is $\frac{2^a}{2^{a-b}}=2^b$.
Let $x,y\in C_j^{(a)}$ for some $j\in\{0,1,\cdots,2^b-1\}$ and $x\neq y$.
We show that $T_a^{(x)}\cap T_a^{(y)}=\phi$.
Now $\abs{T_a^{(x)}\cap T_a^{(y)}}=\abs{x\oplus T_a^{(x)}\cap T_a^{(y)}}=\abs{ T_a\cap T_a^{(x\oplus y)}}$. But $x\oplus y\in S$ and $x\oplus y\neq \mathbf{0}$. By lemma~\ref{pdist3}, $T_a^{(x)}\cap T_a^{(y)}=\phi$.
\end{proof}
The above lemma is used to prove the main theorem of the paper given below.
\begin{theorem}
\label{fracz}

Let $a$ be any non-zero positive integer and $K_a$ be the matrix given by \eqref{dkdasha}. 
Let $B_i$ be a $2^{a-b}\times 2^a$ matrix formed by the rows of $K_a$ indexed by the elements of $C_i^{(a)}$ for $i=0$ to $2^b-1$.

Let $\widetilde{B}_i=HB_i$ where $H$ is a Hadamard matrix of order $2^{a-b}$.
Define 
\begin{eqnarray}
L_a=2^{-\frac{a-b}{2}}
  \left[\begin{array}{c}
  \widetilde{B}_0 \\
  \widetilde{B}_1 \\
   \vdots \\
  \widetilde{B}_{2^b-1}
 \end{array}\right].
\end{eqnarray}
The matrix $L_a$ is a rate-$\frac{a+1}{2^a}$ code with no zero entry for $2^a$ transmit antennas.
\end{theorem}
\begin{proof}
Let 
\begin{equation}
B^\prime=
  \left[\begin{array}{c}
  B_0 \\
  B_1 \\
  \vdots \\
  B_{2^b-1}
 \end{array}\right].
\end{equation}
and 
$\widetilde{H}=I_{2^b}\otimes H$.
The matrix $B^\prime$ is related to $K_a$ by $B^\prime=PK_a$ where $P$ is a permutation matrix of size $2^a\times 2^a$.

We have 
$L_a=2^{-\frac{a-b}{2}}\widetilde{H}B^\prime=2^{-\frac{a-b}{2}}\widetilde{H}PK_a$.
$L_a$ is a CIS-COD as $2^{-\frac{a-b}{2}}\widetilde{H}P$ is a unitary matrix.
By Lemma \ref{kdasha} and Lemma \ref{part}, the number of non-zero elements in any row of $B^\prime$ is $2^b$  and the number of non-zero elements in any row of $L_a$ is $2^{a-b}\cdot 2^b=2^a$. Therefore, all the entries in $L_a$ are non-zero.
\end{proof}
It is clear that $L_a=2^{-\frac{a-b}{2}}\widetilde{H}PK_a=2^{-\frac{a-b}{2}}\widetilde{H}PW_aG_aW_a$. Let $U_a=2^{-\frac{a-b}{2}}\widetilde{H}PW_a$. We have $L_a=U_aG_aW_a$.
For $2^5$ transmit antennas, the pre-multiplying matrix $U_5$ and post-multiplying matrix $W_5$ are displayed in Fig.\ref{udash5} and Fig.\ref{wdash5} respectively.
The code $2L_5$ is displayed in Fig. \ref{ldash5}.

\begin{figure*}[htp]
  \hfill
  \begin{minipage}[t]{.45\textwidth}
    \begin{center}  
      \psfig{file=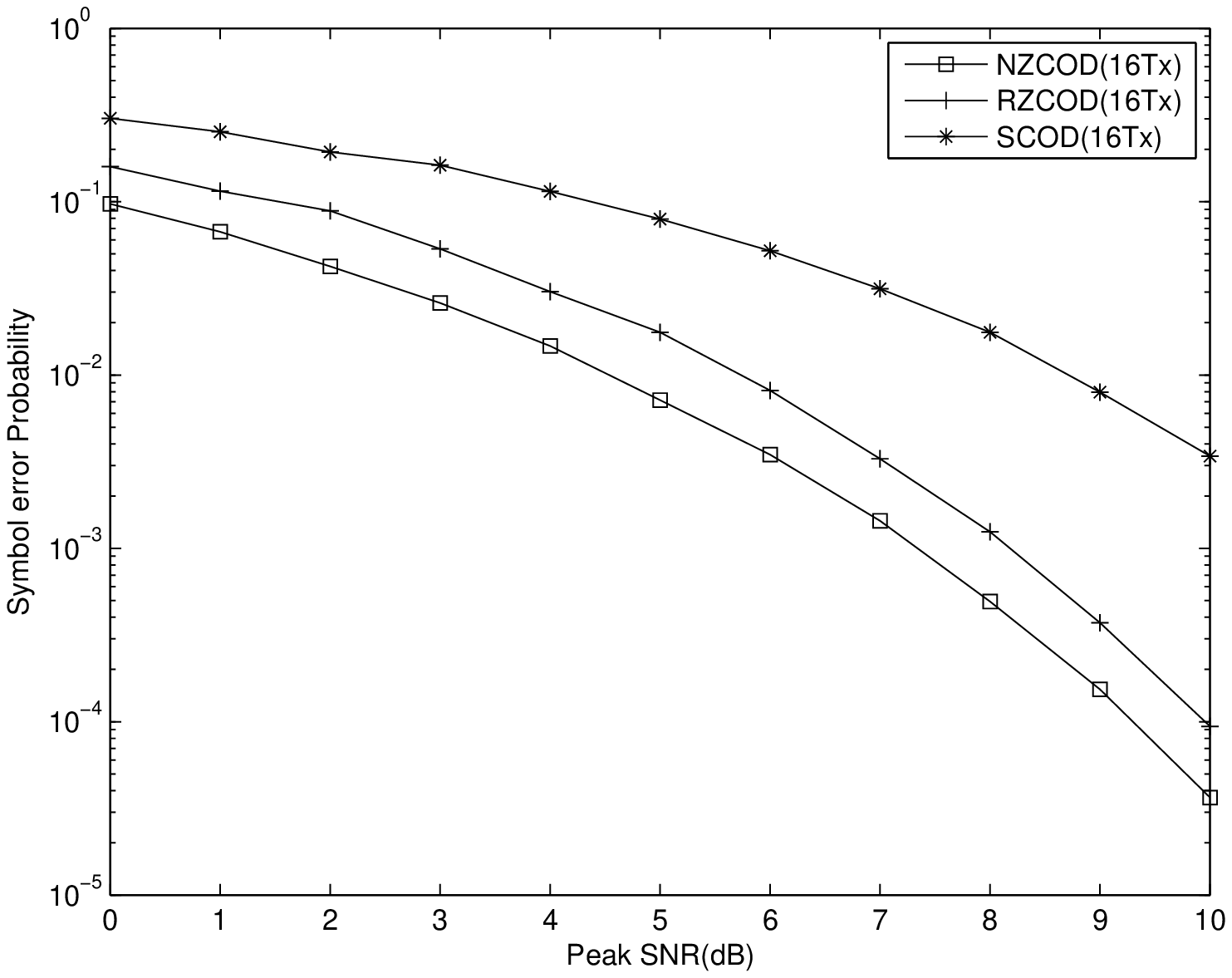, scale=0.5}
\caption{The performance of the NZCOD, RZCOD and SCOD for $16$ Transmit antennas using QAM modulation.} \label{fig1}
    \end{center}
  \end{minipage}
  \hfill
  \begin{minipage}[t]{.45\textwidth}
    \begin{center}  
      \psfig{file=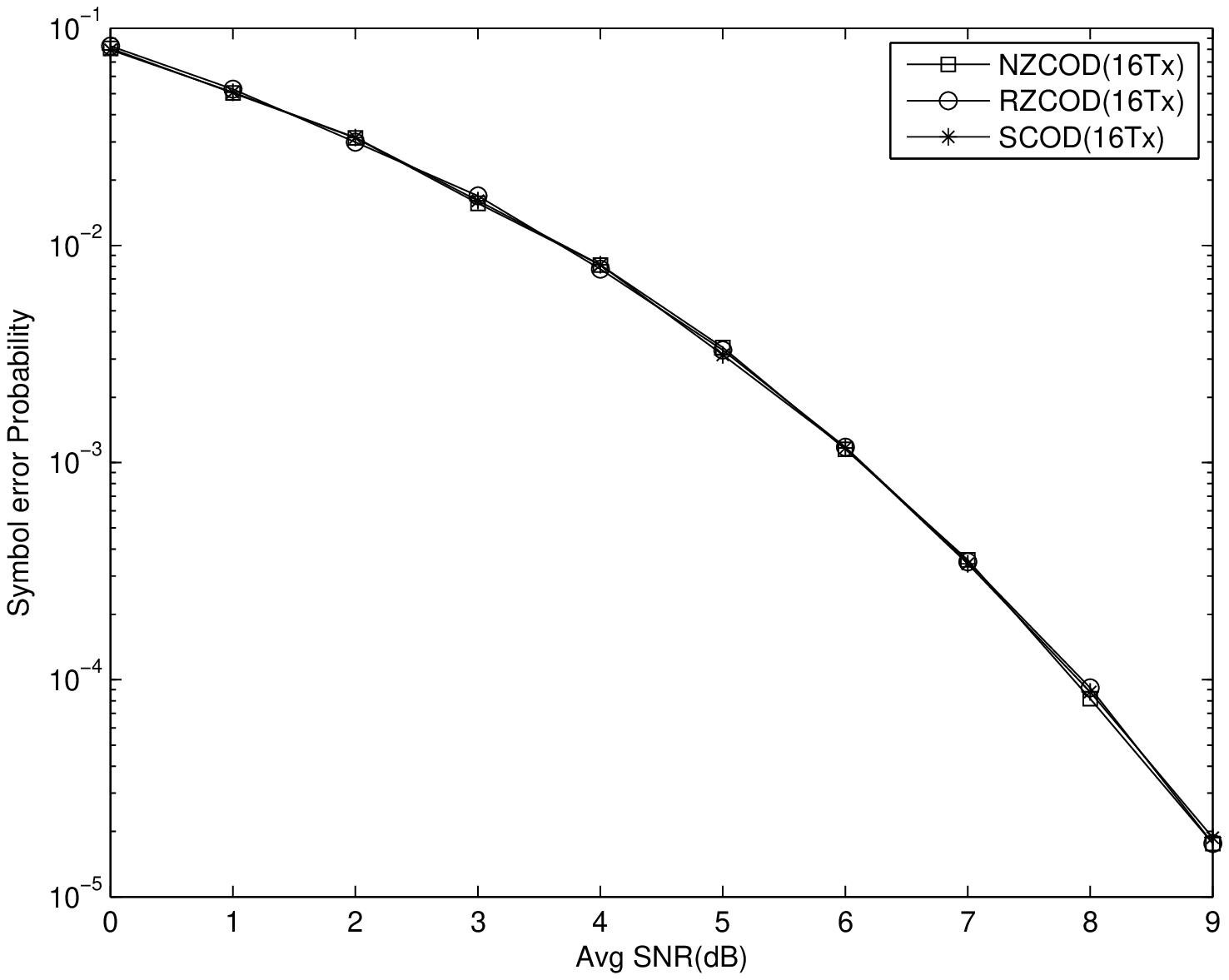, scale=0.5}
\caption{The performance of the NZCOD, RZCOD and SCOD for $16$ Transmit antennas using QAM modulation.} \label{fig2}
\end{center}
\end{minipage}
  \hfill
  \hfill
  \begin{minipage}[t]{.45\textwidth}
    \begin{center}  
      \psfig{file=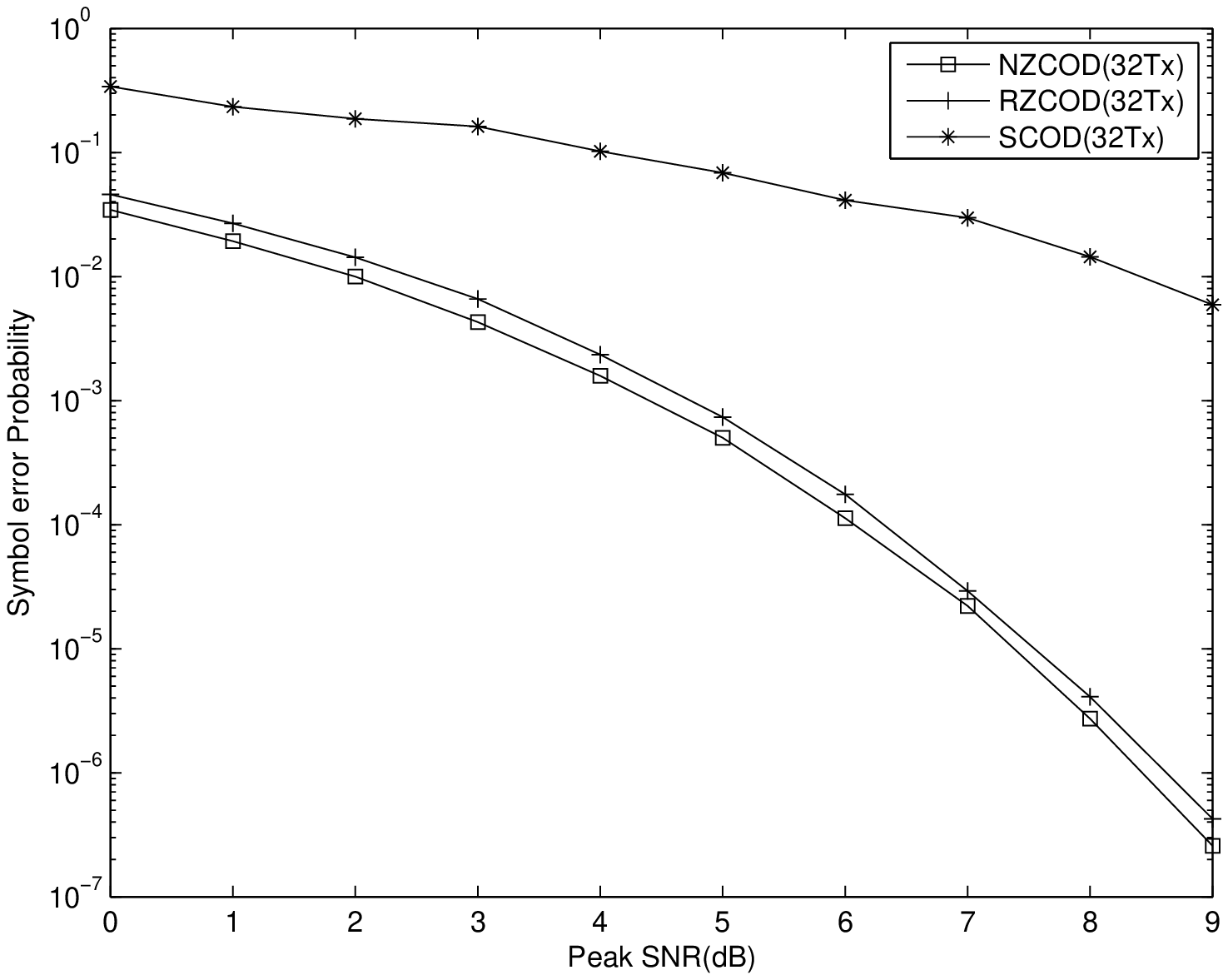, scale=0.5}
\caption{The performance of the NZCOD, RZCOD and SCOD for $32$ Transmit antennas using QAM modulation.} \label{fig1_32}
    \end{center}
  \end{minipage}
  \hfill
  \begin{minipage}[t]{.45\textwidth}
    \begin{center}  
      \psfig{file=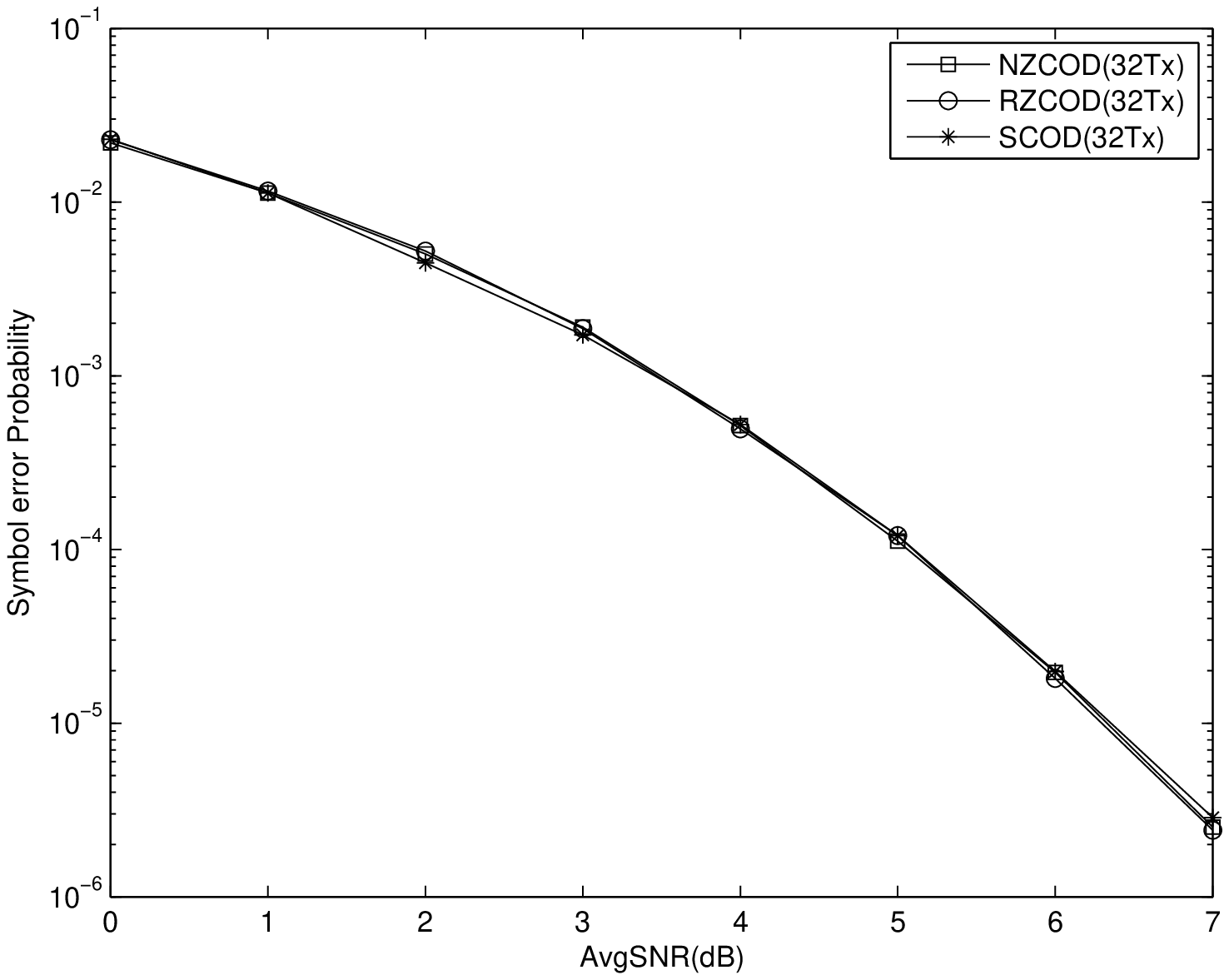, scale=0.5}
\caption{The performance of the NZCOD, RZCOD and SCOD for $32$ Transmit antennas using QAM modulation.} \label{fig2_32}
\end{center}
\end{minipage}
  \hfill

\end{figure*}

{\footnotesize
\begin{figure*}
\begin{equation*}
\hspace{-10pt}
U_5=\frac{1}{2}\left[ 
\begin{array}{ r @{\hspace{4.8pt}} r @{\hspace{6.4pt}} r @{\hspace{6.4pt}} r @{\hspace{6.4pt}} r @{\hspace{6.4pt}} r @{\hspace{6.4pt}} r @{\hspace{6.4pt}} r @{\hspace{6.4pt}} r @{\hspace{6.4pt}} r @{\hspace{6.4pt}} r @{\hspace{6.4pt}} r @{\hspace{6.4pt}} r @{\hspace{6.4pt}} r @{\hspace{6.4pt}} r @{\hspace{6.4pt}} r @{\hspace{6.4pt}} r @{\hspace{6.4pt}} r @{\hspace{6.4pt}} r @{\hspace{6.4pt}} r @{\hspace{6.4pt}} r @{\hspace{6.4pt}} r @{\hspace{6.4pt}} r @{\hspace{6.4pt}} r  @{\hspace{6.4pt}}r @{\hspace{6.4pt}} r @{\hspace{6.4pt}} r @{\hspace{6.4pt}} r @{\hspace{6.4pt}} r @{\hspace{6.4pt}} r @{\hspace{6.4pt}} r @{\hspace{6.4pt}} r } 
 1&  0&  0&  0&  0&  0&  0& 1&  0&  0&  0&  0&  0&  0&  0&  0&  0&  0&  0&  0&  0 &  0&  0&  0&  0&  0& 1&  0&  0& 1&  0&  0\\
 1&  0&  0&  0&  0&  0&  0&-&  0&  0&  0&  0&  0&  0&  0&  0&  0&  0&  0&  0&  0 &  0&  0&  0&  0&  0& 1&  0&  0&-&  0&  0\\
 1&  0&  0&  0&  0&  0&  0& 1&  0&  0&  0&  0&  0&  0&  0&  0&  0&  0&  0&  0&  0 &  0&  0&  0&  0&  0&-&  0&  0&-&  0&  0\\
 1&  0&  0&  0&  0&  0&  0&-&  0&  0&  0&  0&  0&  0&  0&  0&  0&  0&  0&  0&  0&  0&  0&  0&  0&  0&-&  0&  0& 1&  0&  0\\
  0& 1&  0&  0&  0&  0& 1&  0&  0&  0&  0&  0&  0&  0&  0&  0&  0&  0&  0&  0&  0&  0&  0&  0&  0&  0&  0& 1& 1&  0&  0&  0\\
  0& 1&  0&  0&  0&  0&-&  0&  0&  0&  0&  0&  0&  0&  0&  0&  0&  0&  0&  0&  0&  0&  0&  0&  0&  0&  0& 1&-&  0&  0&  0\\
  0& 1&  0&  0&  0&  0& 1&  0&  0&  0&  0&  0&  0&  0&  0&  0&  0&  0&  0&  0&  0&  0&  0&  0&  0&  0&  0&-&-&  0&  0&  0\\
  0& 1&  0&  0&  0&  0&-&  0&  0&  0&  0&  0&  0&  0&  0&  0&  0&  0&  0&  0&  0&  0&  0&  0&  0&  0&  0&-& 1&  0&  0&  0\\
  0&  0& 1&  0&  0& 1&  0&  0&  0&  0&  0&  0&  0&  0&  0&  0&  0&  0&  0&  0&  0 &  0&  0&  0& 1&  0&  0&  0&  0&  0&  0& 1\\
  0&  0& 1&  0&  0&-&  0&  0&  0&  0&  0&  0&  0&  0&  0&  0&  0&  0&  0&  0&  0&  0&  0&  0& 1&  0&  0&  0&  0&  0&  0&-\\
  0&  0& 1&  0&  0& 1&  0&  0&  0&  0&  0&  0&  0&  0&  0&  0&  0&  0&  0&  0&  0&  0&  0&  0&-&  0&  0&  0&  0&  0&  0&-\\
  0&  0& 1&  0&  0&-&  0&  0&  0&  0&  0&  0&  0&  0&  0&  0&  0&  0&  0&  0&  0&  0&  0&  0&-&  0&  0&  0&  0&  0&  0& 1\\
  0&  0&  0& 1& 1&  0&  0&  0&  0&  0&  0&  0&  0&  0&  0&  0&  0&  0&  0&  0&  0&  0&  0&  0&  0& 1&  0&  0&  0&  0& 1&  0\\
  0&  0&  0& 1&-&  0&  0&  0&  0&  0&  0&  0&  0&  0&  0&  0&  0&  0&  0&  0&  0&  0&  0&  0&  0& 1&  0&  0&  0&  0&-&  0\\
  0&  0&  0& 1& 1&  0&  0&  0&  0&  0&  0&  0&  0&  0&  0&  0&  0&  0&  0&  0&  0&  0&  0&  0&  0&-&  0&  0&  0&  0&-&  0\\
  0&  0&  0& 1&-&  0&  0&  0&  0&  0&  0&  0&  0&  0&  0&  0&  0&  0&  0&  0&  0&  0&  0&  0&  0&-&  0&  0&  0&  0& 1&  0\\
  0&  0&  0&  0&  0&  0&  0&  0&  s&  s&  0&  0&  0&  0&  s& -s&  0&  0&  s&  s&  s& -s&  0&  0&  0&  0&  0&  0&  0&  0&  0&  0\\
  0&  0&  0&  0&  0&  0&  0&  0&  s&  s&  0&  0&  0&  0& -s&  s&  0&  0&  s&  s& -s&  s&  0&  0&  0&  0&  0&  0&  0&  0&  0&  0\\
  0&  0&  0&  0&  0&  0&  0&  0&  s&  s&  0&  0&  0&  0&  s& -s&  0&  0& -s& -s& -s&  s&  0&  0&  0&  0&  0&  0&  0&  0&  0&  0\\
  0&  0&  0&  0&  0&  0&  0&  0&  s&  s&  0&  0&  0&  0& -s&  s&  0&  0& -s& -s&  s& -s&  0&  0&  0&  0&  0&  0&  0&  0&  0&  0\\
  0&  0&  0&  0&  0&  0&  0&  0&  s& -s&  0&  0&  0&  0&  s&  s&  0&  0&  s& -s&  s&  s&  0&  0&  0&  0&  0&  0&  0&  0&  0&  0\\
  0&  0&  0&  0&  0&  0&  0&  0&  s& -s&  0&  0&  0&  0& -s& -s&  0&  0&  s& -s& -s& -s&  0&  0&  0&  0&  0&  0&  0&  0&  0&  0\\
  0&  0&  0&  0&  0&  0&  0&  0&  s& -s&  0&  0&  0&  0&  s&  s&  0&  0& -s&  s& -s& -s&  0&  0&  0&  0&  0&  0&  0&  0&  0&  0\\
  0&  0&  0&  0&  0&  0&  0&  0&  s& -s&  0&  0&  0&  0& -s& -s&  0&  0& -s&  s&  s&  s&  0&  0&  0&  0&  0&  0&  0&  0&  0&  0\\
  0&  0&  0&  0&  0&  0&  0&  0&  0&  0&  s&  s&  s& -s&  0&  0&  s&  s&  0&  0&  0&  0&  s& -s&  0&  0&  0&  0&  0&  0&  0&  0\\
  0&  0&  0&  0&  0&  0&  0&  0&  0&  0&  s&  s& -s&  s&  0&  0&  s&  s&  0&  0&  0&  0& -s&  s&  0&  0&  0&  0&  0&  0&  0&  0\\
  0&  0&  0&  0&  0&  0&  0&  0&  0&  0&  s&  s&  s& -s&  0&  0& -s& -s&  0&  0&  0&  0& -s&  s&  0&  0&  0&  0&  0&  0&  0&  0\\
  0&  0&  0&  0&  0&  0&  0&  0&  0&  0&  s&  s& -s&  s&  0&  0& -s& -s&  0&  0&  0&  0&  s& -s&  0&  0&  0&  0&  0&  0&  0&  0\\
  0&  0&  0&  0&  0&  0&  0&  0&  0&  0&  s& -s&  s&  s&  0&  0&  s& -s&  0&  0&  0&  0&  s&  s&  0&  0&  0&  0&  0&  0&  0&  0\\
  0&  0&  0&  0&  0&  0&  0&  0&  0&  0&  s& -s& -s& -s&  0&  0&  s& -s&  0&  0&  0&  0& -s& -s&  0&  0&  0&  0&  0&  0&  0&  0\\
  0&  0&  0&  0&  0&  0&  0&  0&  0&  0&  s& -s&  s&  s&  0&  0& -s&  s&  0&  0&  0&  0& -s& -s&  0&  0&  0&  0&  0&  0&  0&  0\\
  0&  0&  0&  0&  0&  0&  0&  0&  0&  0&  s& -s& -s& -s&  0&  0& -s&  s&  0&  0&  0&  0&  s&  s&  0&  0&  0&  0&  0&  0&  0&  0
\end{array}
\right]
\end{equation*}
 \caption{The pre-multiplying matrix $U_5$ for $32$ antennas where $s=\frac{1}{\sqrt{2}}$}
\label{udash5}

\end{figure*}

{\footnotesize
\begin{figure*}
\begin{equation*}
\hspace{-10pt}
W_5=\left[ 
\begin{array}{ r @{\hspace{4.8pt}} r @{\hspace{6.4pt}} r @{\hspace{6.4pt}} r @{\hspace{6.4pt}} r @{\hspace{6.4pt}} r @{\hspace{6.4pt}} r @{\hspace{6.4pt}} r @{\hspace{6.4pt}} r @{\hspace{6.4pt}} r @{\hspace{6.4pt}} r @{\hspace{6.4pt}} r @{\hspace{6.4pt}} r @{\hspace{6.4pt}} r @{\hspace{6.4pt}} r @{\hspace{6.4pt}} r @{\hspace{6.4pt}} r @{\hspace{6.4pt}} r @{\hspace{6.4pt}} r @{\hspace{6.4pt}} r @{\hspace{6.4pt}} r @{\hspace{6.4pt}} r @{\hspace{6.4pt}} r @{\hspace{6.4pt}} r  @{\hspace{6.4pt}}r @{\hspace{6.4pt}} r @{\hspace{6.4pt}} r @{\hspace{6.4pt}} r @{\hspace{6.4pt}} r @{\hspace{6.4pt}} r @{\hspace{6.4pt}} r @{\hspace{6.4pt}} r } 
 1&  0&  0&  0&  0&  0&  0&  0&  0&  0&  0&  0&  0&  0&  0&  0&  0&  0&  0&  0&  0 &  0&  0&    0&  0&  0&  0&  0&  0&  0&  0&  0 \\          
  0& 1&  0&  0&  0&  0&  0&  0&  0&  0&  0&  0&  0&  0&  0&  0&  0&  0&  0&  0&  0&  0&  0&     0&  0&  0&  0&  0&  0&  0&  0&  0 \\
  0&  0& 1&  0&  0&  0&  0&  0&  0&  0&  0&  0&  0&  0&  0&  0&  0&  0&  0&  0&  0 &  0&  0&  0&  0&  0&  0&  0&  0&  0&  0&  0 \\
  0&  0&  0& 1&  0&  0&  0&  0&  0&  0&  0&  0&  0&  0&  0&  0&  0&  0&  0&  0&  0 &  0&  0&  0&  0&  0&  0&  0&  0&  0&  0&  0 \\
  0&  0&  0&  0& 1&  0&  0&  0&  0&  0&  0&  0&  0&  0&  0&  0&  0&  0&  0&  0&  0 &  0&  0&  0&  0&  0&  0&  0&  0&  0&  0&  0 \\
  0&  0&  0&  0&  0& 1&  0&  0&  0&  0&  0&  0&  0&  0&  0&  0&  0&  0&  0&  0&  0 &  0&  0&  0&  0&  0&  0&  0&  0&  0&  0&  0 \\
  0&  0&  0&  0&  0&  0& 1&  0&  0&  0&  0&  0&  0&  0&  0&  0&  0&  0&  0&  0&  0 &  0&  0&  0&  0&  0&  0&  0&  0&  0&  0&  0 \\
  0&  0&  0&  0&  0&  0&  0& 1&  0&  0&  0&  0&  0&  0&  0&  0&  0&  0&  0&  0&  0 &  0&  0&  0&  0&  0&  0&  0&  0&  0&  0&  0 \\
  0&  0&  0&  0&  0&  0&  0&  0&  s&  s&  0&  0&  0&  0&  0&  0&  0&  0&  0&  0&  0 &  0&  0&  0&  0&  0&  0&  0&  0&  0&  0&  0 \\
  0&  0&  0&  0&  0&  0&  0&  0&  s& -s&  0&  0&  0&  0&  0&  0&  0&  0&  0&  0&  0 &  0&  0&  0&  0&  0&  0&  0&  0&  0&  0&  0 \\
  0&  0&  0&  0&  0&  0&  0&  0&  0&  0&  s&  s&  0&  0&  0&  0&  0&  0&  0&  0&  0 &  0&  0&  0&  0&  0&  0&  0&  0&  0&  0&  0 \\
  0&  0&  0&  0&  0&  0&  0&  0&  0&  0&  s& -s&  0&  0&  0&  0&  0&  0&  0&  0&  0 &  0&  0&  0&  0&  0&  0&  0&  0&  0&  0&  0 \\
  0&  0&  0&  0&  0&  0&  0&  0&  0&  0&  0&  0&  s&  s&  0&  0&  0&  0&  0&  0&  0 &  0&  0&  0&  0&  0&  0&  0&  0&  0&  0&  0 \\
  0&  0&  0&  0&  0&  0&  0&  0&  0&  0&  0&  0&  s& -s&  0&  0&  0&  0&  0&  0&  0&  0&  0&  0&  0&  0&  0&  0&  0&  0&  0&  0 \\
  0&  0&  0&  0&  0&  0&  0&  0&  0&  0&  0&  0&  0&  0&  s&  s&  0&  0&  0&  0&  0 &  0&  0&  0&  0&  0&  0&  0&  0&  0&  0&  0 \\
  0&  0&  0&  0&  0&  0&  0&  0&  0&  0&  0&  0&  0&  0&  s& -s&  0&  0&  0&  0&  0 &  0&  0&  0&  0&  0&  0&  0&  0&  0&  0&  0 \\
  0&  0&  0&  0&  0&  0&  0&  0&  0&  0&  0&  0&  0&  0&  0&  0&  s&  s&  0&  0&  0&  0&  0&  0&  0&  0&  0&  0&  0&  0&  0&  0 \\
  0&  0&  0&  0&  0&  0&  0&  0&  0&  0&  0&  0&  0&  0&  0&  0&  s& -s&  0&  0&  0&  0&  0&  0&  0&  0&  0&  0&  0&  0&  0&  0 \\
  0&  0&  0&  0&  0&  0&  0&  0&  0&  0&  0&  0&  0&  0&  0&  0&  0&  0&  s&  s&  0&  0&  0&  0&  0&  0&  0&  0&  0&  0&  0&  0 \\
  0&  0&  0&  0&  0&  0&  0&  0&  0&  0&  0&  0&  0&  0&  0&  0&  0&  0&  s& -s&  0&  0&  0&  0&  0&  0&  0&  0&  0&  0&  0&  0 \\
  0&  0&  0&  0&  0&  0&  0&  0&  0&  0&  0&  0&  0&  0&  0&  0&  0&  0&  0&  0&  s &  s&  0&  0&  0&  0&  0&  0&  0&  0&  0&  0\\
  0&  0&  0&  0&  0&  0&  0&  0&  0&  0&  0&  0&  0&  0&  0&  0&  0&  0&  0&  0&  s & -s&  0&  0&  0&  0&  0&  0&  0&  0&  0&  0\\
  0&  0&  0&  0&  0&  0&  0&  0&  0&  0&  0&  0&  0&  0&  0&  0&  0&  0&  0&  0&  0&  0&  s&  s&  0&  0&  0&  0&  0&  0&  0&  0\\
  0&  0&  0&  0&  0&  0&  0&  0&  0&  0&  0&  0&  0&  0&  0&  0&  0&  0&  0&  0&  0 &  0&  s& -s&  0&  0&  0&  0&  0&  0&  0&  0\\
  0&  0&  0&  0&  0&  0&  0&  0&  0&  0&  0&  0&  0&  0&  0&  0&  0&  0&  0&  0&  0
   &  0&  0&  0& 1&  0&  0&  0&  0&  0&  0&  0\\
  0&  0&  0&  0&  0&  0&  0&  0&  0&  0&  0&  0&  0&  0&  0&  0&  0&  0&  0&  0&  0&  0&  0&  0& 0& 1&    0&  0&  0&  0&  0&  0\\
  0&  0&  0&  0&  0&  0&  0&  0&  0&  0&  0&  0&  0&  0&  0&  0&  0&  0&  0&  0&  0&  0&  0&  0& 0&  0& 1&    0&  0&  0&  0&  0\\
  0&  0&  0&  0&  0&  0&  0&  0&  0&  0&  0&  0&  0&  0&  0&  0&  0&  0&  0&  0&  0&  0&  0&  0& 0&  0&  0& 1&    0&  0&  0&  0\\
  0&  0&  0&  0&  0&  0&  0&  0&  0&  0&  0&  0&  0&  0&  0&  0&  0&  0&  0&  0&  0&  0&  0&  0& 0&  0&  0&  0& 1&    0&  0&  0\\
  0&  0&  0&  0&  0&  0&  0&  0&  0&  0&  0&  0&  0&  0&  0&  0&  0&  0&  0&  0&  0&  0&  0&  0 &0&  0&  0&  0&  0& 1&    0&  0\\
  0&  0&  0&  0&  0&  0&  0&  0&  0&  0&  0&  0&  0&  0&  0&  0&  0&  0&  0&  0&  0&  0&  0&  0& 0&  0&  0&  0&  0&  0& 1&    0\\
  0&  0&  0&  0&  0&  0&  0&  0&  0&  0&  0&  0&  0&  0&  0&  0&  0&  0&  0&  0&  0&  0&  0&  0 &  0&  0&  0&  0&  0&  0&  0& 1
\end{array}
\right]
\end{equation*}
 \caption{The post-multiplying matrix $W_5$ for $32$ antennas where $s=\frac{1}{\sqrt{2}}$}
\label{wdash5}
 \end{figure*}

\begin{figure*}
{\footnotesize
\begin{equation*}
\hspace{-40pt}
\left [ \hspace{-5pt}
\begin{array}{ r @{\hspace{.8pt}} r @{\hspace{.8pt}} r @{\hspace{.8pt}} r @{\hspace{.8pt}} r @{\hspace{.8pt}} r @{\hspace{.8pt}} r @{\hspace{.8pt}} r @{\hspace{.8pt}} r @{\hspace{.8pt}} r @{\hspace{.8pt}} r @{\hspace{.8pt}} r @{\hspace{.8pt}} r @{\hspace{.8pt}} r @{\hspace{.8pt}} r @{\hspace{.8pt}} r @{\hspace{.8pt}} r @{\hspace{.8pt}} r @{\hspace{.8pt}} r @{\hspace{.8pt}} r @{\hspace{.8pt}} r @{\hspace{.8pt}} r @{\hspace{.8pt}} r @{\hspace{.8pt}} r  @{\hspace{.8pt}}r @{\hspace{.8pt}} r @{\hspace{.8pt}} r @{\hspace{.8pt}} r @{\hspace{.8pt}} r @{\hspace{.8pt}} r @{\hspace{.8pt}} r @{\hspace{.8pt}} r }
x_1&     -x_2^*&     -x_3^*&  x_4&     -x_4^*& -x_3&  x_2&      x_1^*&   -y_5^*&   -y_5^*& y_6& y_6& y_6&-y_6&   -y_5^*&    y_5^*&   -y_6^*&   -y_6^*&-y_5&-y_5&-y_5& y_5&   -y_6^*&    y_6^*&  x_3&  x_4&      x_1^*&      x_2^*& -x_2&  x_1&     -x_4^*&      x_3^*\\
 x_1&     -x_2^*&     -x_3^*& -x_4&     -x_4^*&  x_3& -x_2&     -x_1^*&   -y_5^*&   -y_5^*& y_6& y_6&-y_6& y_6&    y_5^*&   -y_5^*&   -y_6^*&   -y_6^*&-y_5&-y_5& y_5&-y_5&    y_6^*&   -y_6^*&  x_3& -x_4&      x_1^*&      x_2^*&  x_2& -x_1&     -x_4^*&     -x_3^*\\
 x_1&     -x_2^*&     -x_3^*&  x_4&     -x_4^*& -x_3&  x_2&      x_1^*&   -y_5^*&   -y_5^*&-y_6&-y_6&-y_6& y_6&   -y_5^*&    y_5^*&   -y_6^*&   -y_6^*& y_5& y_5& y_5&-y_5&   -y_6^*&    y_6^*& -x_3& -x_4&     -x_1^*&     -x_2^*&  x_2& -x_1&      x_4^*&     -x_3^*\\
 x_1&     -x_2^*&     -x_3^*& -x_4&     -x_4^*&  x_3& -x_2&     -x_1^*&   -y_5^*&   -y_5^*&-y_6&-y_6& y_6&-y_6&    y_5^*&   -y_5^*&   -y_6^*&   -y_6^*& y_5& y_5&-y_5& y_5&    y_6^*&   -y_6^*& -x_3&  x_4&     -x_1^*&     -x_2^*& -x_2&  x_1&      x_4^*&      x_3^*\\
 x_2&      x_1^*&  x_4&     -x_3^*& -x_3&     -x_4^*&  x_1&     -x_2^*&   -y_5^*&    y_5^*& y_6&-y_6& y_6& y_6&   -y_5^*&   -y_5^*&   -y_6^*&    y_6^*&-y_5& y_5&-y_5&-y_5&   -y_6^*&   -y_6^*&  x_4&  x_3& -x_2&  x_1&      x_1^*&      x_2^*&      x_3^*&     -x_4^*\\
 x_2&      x_1^*& -x_4&     -x_3^*&  x_3&     -x_4^*& -x_1&      x_2^*&   -y_5^*&    y_5^*& y_6&-y_6&-y_6&-y_6&    y_5^*&    y_5^*&   -y_6^*&    y_6^*&-y_5& y_5& y_5& y_5&    y_6^*&    y_6^*& -x_4&  x_3& -x_2&  x_1&     -x_1^*&     -x_2^*&     -x_3^*&     -x_4^*\\
 x_2&      x_1^*&  x_4&     -x_3^*& -x_3&     -x_4^*&  x_1&     -x_2^*&   -y_5^*&    y_5^*&-y_6& y_6&-y_6&-y_6&   -y_5^*&   -y_5^*&   -y_6^*&    y_6^*& y_5&-y_5& y_5& y_5&   -y_6^*&   -y_6^*& -x_4& -x_3&  x_2& -x_1&     -x_1^*&     -x_2^*&     -x_3^*&      x_4^*\\
 x_2&      x_1^*& -x_4&     -x_3^*&  x_3&     -x_4^*& -x_1&      x_2^*&   -y_5^*&    y_5^*&-y_6& y_6& y_6& y_6&    y_5^*&    y_5^*&   -y_6^*&    y_6^*& y_5&-y_5&-y_5&-y_5&    y_6^*&    y_6^*&  x_4& -x_3&  x_2& -x_1&      x_1^*&      x_2^*&      x_3^*&      x_4^*\\
 x_3&  x_4&      x_1^*&      x_2^*& -x_2&  x_1&     -x_4^*&      x_3^*& y_6& y_6&   -y_5^*&   -y_5^*&   -y_5^*&    y_5^*& y_6&-y_6&-y_5&-y_5&   -y_6^*&   -y_6^*&   -y_6^*&    y_6^*&-y_5& y_5&  x_1&     -x_2^*&     -x_3^*&  x_4&     -x_4^*& -x_3&  x_2&      x_1^*\\
 x_3& -x_4&      x_1^*&      x_2^*&  x_2& -x_1&     -x_4^*&     -x_3^*& y_6& y_6&   -y_5^*&   -y_5^*&    y_5^*&   -y_5^*&-y_6& y_6&-y_5&-y_5&   -y_6^*&   -y_6^*&    y_6^*&   -y_6^*& y_5&-y_5&  x_1&     -x_2^*&     -x_3^*& -x_4&     -x_4^*&  x_3& -x_2&     -x_1^*\\
 x_3&  x_4&      x_1^*&      x_2^*& -x_2&  x_1&     -x_4^*&      x_3^*&-y_6&-y_6&   -y_5^*&   -y_5^*&   -y_5^*&    y_5^*&-y_6& y_6& y_5& y_5&   -y_6^*&   -y_6^*&   -y_6^*&    y_6^*& y_5&-y_5& -x_1&      x_2^*&      x_3^*& -x_4&      x_4^*&  x_3& -x_2&     -x_1^*\\
 x_3& -x_4&      x_1^*&      x_2^*&  x_2& -x_1&     -x_4^*&     -x_3^*&-y_6&-y_6&   -y_5^*&   -y_5^*&    y_5^*&   -y_5^*& y_6&-y_6& y_5& y_5&   -y_6^*&   -y_6^*&    y_6^*&   -y_6^*&-y_5& y_5& -x_1&      x_2^*&      x_3^*&  x_4&      x_4^*& -x_3&  x_2&      x_1^*\\
 x_4&  x_3& -x_2&  x_1&      x_1^*&      x_2^*&      x_3^*&     -x_4^*& y_6&-y_6&   -y_5^*&    y_5^*&   -y_5^*&   -y_5^*& y_6& y_6&-y_5& y_5&   -y_6^*&    y_6^*&   -y_6^*&   -y_6^*&-y_5&-y_5&  x_2&      x_1^*&  x_4&     -x_3^*& -x_3&     -x_4^*&  x_1&     -x_2^*\\
-x_4&  x_3& -x_2&  x_1&     -x_1^*&     -x_2^*&     -x_3^*&     -x_4^*& y_6&-y_6&   -y_5^*&    y_5^*&    y_5^*&    y_5^*&-y_6&-y_6&-y_5& y_5&   -y_6^*&    y_6^*&    y_6^*&    y_6^*& y_5& y_5&  x_2&      x_1^*& -x_4&     -x_3^*&  x_3&     -x_4^*& -x_1&      x_2^*\\
 x_4&  x_3& -x_2&  x_1&      x_1^*&      x_2^*&      x_3^*&     -x_4^*&-y_6& y_6&   -y_5^*&    y_5^*&   -y_5^*&   -y_5^*&-y_6&-y_6& y_5&-y_5&   -y_6^*&    y_6^*&   -y_6^*&   -y_6^*& y_5& y_5& -x_2&     -x_1^*& -x_4&      x_3^*&  x_3&      x_4^*& -x_1&      x_2^*\\
-x_4&  x_3& -x_2&  x_1&     -x_1^*&     -x_2^*&     -x_3^*&     -x_4^*&-y_6& y_6&   -y_5^*&    y_5^*&    y_5^*&    y_5^*& y_6& y_6& y_5&-y_5&   -y_6^*&    y_6^*&    y_6^*&    y_6^*&-y_5&-y_5& -x_2&     -x_1^*&  x_4&      x_3^*& -x_3&      x_4^*&  x_1&     -x_2^*\\
y_5& y_5& y_6& y_6& y_6&-y_6& y_5&-y_5&\tilde{x}_1^*&-\tilde{x}_2&      x_3^*& -x_4&      x_4^*&  x_3&  \tilde{x}_2^*&  \tilde{x}_1& -x_3& -x_4&  \tilde{x}_1& \tilde{x}_2& -\tilde{x}_2^*&\tilde{x}_1^*&      x_4^*&     -x_3^*&   -y_6^*&   -y_6^*&    y_5^*&    y_5^*&    y_5^*&   -y_5^*&   -y_6^*&    y_6^*\\
y_5& y_5& y_6& y_6&-y_6& y_6&-y_5& y_5&\tilde{x}_1^*&-\tilde{x}_2&      x_3^*&  x_4&      x_4^*& -x_3& -\tilde{x}_2^*& -\tilde{x}_1& -x_3&  x_4&  \tilde{x}_1& \tilde{x}_2&  \tilde{x}_2^*& -\tilde{x}_1^*&      x_4^*&      x_3^*&   -y_6^*&   -y_6^*&    y_5^*&    y_5^*&   -y_5^*&    y_5^*&    y_6^*&   -y_6^*\\
y_5& y_5&-y_6&-y_6&-y_6& y_6& y_5&-y_5&\tilde{x}_1^*&-\tilde{x}_2&      x_3^*& -x_4&      x_4^*&  x_3&  \tilde{x}_2^*&  \tilde{x}_1&  x_3&  x_4& -\tilde{x}_1&-\tilde{x}_2&  \tilde{x}_2^*& -\tilde{x}_1^*&     -x_4^*&      x_3^*&   -y_6^*&   -y_6^*&   -y_5^*&   -y_5^*&   -y_5^*&    y_5^*&   -y_6^*&    y_6^*\\
y_5& y_5&-y_6&-y_6& y_6&-y_6&-y_5& y_5&\tilde{x}_1^*&-\tilde{x}_2&      x_3^*&  x_4&      x_4^*& -x_3& -\tilde{x}_2^*& -\tilde{x}_1&  x_3& -x_4& -\tilde{x}_1&-\tilde{x}_2& -\tilde{x}_2^*&\tilde{x}_1^*&     -x_4^*&     -x_3^*&   -y_6^*&   -y_6^*&   -y_5^*&   -y_5^*&    y_5^*&   -y_5^*&    y_6^*&   -y_6^*\\
y_5&-y_5& y_6&-y_6& y_6& y_6& y_5& y_5&  \tilde{x}_2^*&  \tilde{x}_1& -x_4&      x_3^*&  x_3&      x_4^*&\tilde{x}_1^*&-\tilde{x}_2& -x_4& -x_3& -\tilde{x}_2^*&\tilde{x}_1^*&  \tilde{x}_1& \tilde{x}_2&     -x_3^*&      x_4^*&   -y_6^*&    y_6^*&    y_5^*&   -y_5^*&    y_5^*&    y_5^*&   -y_6^*&   -y_6^*\\
y_5&-y_5& y_6&-y_6&-y_6&-y_6&-y_5&-y_5&  \tilde{x}_2^*&  \tilde{x}_1&  x_4&      x_3^*& -x_3&      x_4^*& -\tilde{x}_1^*& \tilde{x}_2&  x_4& -x_3& -\tilde{x}_2^*&\tilde{x}_1^*& -\tilde{x}_1&-\tilde{x}_2&      x_3^*&      x_4^*&   -y_6^*&    y_6^*&    y_5^*&   -y_5^*&   -y_5^*&   -y_5^*&    y_6^*&    y_6^*\\
y_5&-y_5&-y_6& y_6&-y_6&-y_6& y_5& y_5&  \tilde{x}_2^*&  \tilde{x}_1& -x_4&      x_3^*&  x_3&      x_4^*&\tilde{x}_1^*&-\tilde{x}_2&  x_4&  x_3&  \tilde{x}_2^*& -\tilde{x}_1^*& -\tilde{x}_1&-\tilde{x}_2&      x_3^*&     -x_4^*&   -y_6^*&    y_6^*&   -y_5^*&    y_5^*&   -y_5^*&   -y_5^*&   -y_6^*&   -y_6^*\\
y_5&-y_5&-y_6& y_6& y_6& y_6&-y_5&-y_5&  \tilde{x}_2^*&  \tilde{x}_1&  x_4&      x_3^*& -x_3&      x_4^*& -\tilde{x}_1^*& \tilde{x}_2& -x_4&  x_3&  \tilde{x}_2^*& -\tilde{x}_1^*&  \tilde{x}_1& \tilde{x}_2&     -x_3^*&     -x_4^*&   -y_6^*&    y_6^*&   -y_5^*&    y_5^*&    y_5^*&    y_5^*&    y_6^*&    y_6^*\\
y_6& y_6& y_5& y_5& y_5&-y_5& y_6&-y_6& -x_3& -x_4&  \tilde{x}_1& \tilde{x}_2& -\tilde{x}_2^*&\tilde{x}_1^*&      x_4^*&     -x_3^*&\tilde{x}_1^*&-\tilde{x}_2&      x_3^*& -x_4&      x_4^*&  x_3&  \tilde{x}_2^*&  \tilde{x}_1&    y_5^*&    y_5^*&   -y_6^*&   -y_6^*&   -y_6^*&    y_6^*&    y_5^*&   -y_5^*\\
y_6& y_6& y_5& y_5&-y_5& y_5&-y_6& y_6& -x_3&  x_4&  \tilde{x}_1& \tilde{x}_2&  \tilde{x}_2^*& -\tilde{x}_1^*&      x_4^*&      x_3^*&\tilde{x}_1^*&-\tilde{x}_2&      x_3^*&  x_4&      x_4^*& -x_3& -\tilde{x}_2^*& -\tilde{x}_1&    y_5^*&    y_5^*&   -y_6^*&   -y_6^*&    y_6^*&   -y_6^*&   -y_5^*&    y_5^*\\
-y_6&-y_6& y_5& y_5& y_5&-y_5&-y_6& y_6& -x_3& -x_4&  \tilde{x}_1& \tilde{x}_2& -\tilde{x}_2^*&\tilde{x}_1^*&      x_4^*&     -x_3^*& -\tilde{x}_1^*& \tilde{x}_2&     -x_3^*&  x_4&     -x_4^*& -x_3& -\tilde{x}_2^*& -\tilde{x}_1&   -y_5^*&   -y_5^*&   -y_6^*&   -y_6^*&   -y_6^*&    y_6^*&   -y_5^*&    y_5^*\\
-y_6&-y_6& y_5& y_5&-y_5& y_5& y_6&-y_6& -x_3&  x_4&  \tilde{x}_1& \tilde{x}_2&  \tilde{x}_2^*& -\tilde{x}_1^*&      x_4^*&      x_3^*& -\tilde{x}_1^*& \tilde{x}_2&     -x_3^*& -x_4&     -x_4^*&  x_3&  \tilde{x}_2^*&  \tilde{x}_1&   -y_5^*&   -y_5^*&   -y_6^*&   -y_6^*&    y_6^*&   -y_6^*&    y_5^*&   -y_5^*\\
y_6&-y_6& y_5&-y_5& y_5& y_5& y_6& y_6& -x_4& -x_3& -\tilde{x}_2^*&\tilde{x}_1^*&  \tilde{x}_1& \tilde{x}_2&     -x_3^*&      x_4^*&  \tilde{x}_2^*&  \tilde{x}_1& -x_4&      x_3^*&  x_3&      x_4^*&\tilde{x}_1^*&-\tilde{x}_2&    y_5^*&   -y_5^*&   -y_6^*&    y_6^*&   -y_6^*&   -y_6^*&    y_5^*&    y_5^*\\
y_6&-y_6& y_5&-y_5&-y_5&-y_5&-y_6&-y_6&  x_4& -x_3& -\tilde{x}_2^*&\tilde{x}_1^*& -\tilde{x}_1&-\tilde{x}_2&      x_3^*&      x_4^*&  \tilde{x}_2^*&  \tilde{x}_1&  x_4&      x_3^*& -x_3&      x_4^*& -\tilde{x}_1^*& \tilde{x}_2&    y_5^*&   -y_5^*&   -y_6^*&    y_6^*&    y_6^*&    y_6^*&   -y_5^*&   -y_5^*\\
-y_6& y_6& y_5&-y_5& y_5& y_5&-y_6&-y_6& -x_4& -x_3& -\tilde{x}_2^*&\tilde{x}_1^*&  \tilde{x}_1& \tilde{x}_2&     -x_3^*&      x_4^*& -\tilde{x}_2^*& -\tilde{x}_1&  x_4&     -x_3^*& -x_3&     -x_4^*& -\tilde{x}_1^*& \tilde{x}_2&   -y_5^*&    y_5^*&   -y_6^*&    y_6^*&   -y_6^*&   -y_6^*&   -y_5^*&   -y_5^*\\
-y_6& y_6& y_5&-y_5&-y_5&-y_5& y_6& y_6&  x_4& -x_3& -\tilde{x}_2^*&\tilde{x}_1^*& -\tilde{x}_1&-\tilde{x}_2&      x_3^*&      x_4^*& -\tilde{x}_2^*& -\tilde{x}_1& -x_4&     -x_3^*&  x_3&     -x_4^*&\tilde{x}_1^*&-\tilde{x}_2&   -y_5^*&    y_5^*&   -y_6^*&    y_6^*&    y_6^*&    y_6^*&    y_5^*&    y_5^*\\
\end{array}
\right]
\end{equation*}
}

\caption{The $[32,32,6]$ code $L_5$ with no zero entry where $y_5=\frac{x_5}{\sqrt{2}},y_6=\frac{x_6}{\sqrt{2}}, \tilde{x}_1=x_{1,2}$ and $\tilde{x}_2=x_{2,1}$}
\label{ldash5}
\end{figure*}
\subsection{Signaling Complexity of $L_a$}
\label{subsec2_1}
Notice that from the construction of the code $L_a,$ only two coordinate interleaved variables, namely, variables $x_{1,2}$ and $x_{2,1},$ appear, irrespective of the value of $a.$ The other variables appear either as they are or with conjugation and possible multiplication by -1. This means that the increase in the signaling complexity is only marginal compared to a COD (if at all it existed!!) with no zero-entry and also all the variables appearing without any nontrivial linear combinations including coordinate interleaving.    

\section{Simulation Results}
\label{sec3}
The symbol error rate performance of the code with no zero entry constructed in this paper (denoted as NZCOD in the plots which means COD with No Zero) for $16$ antennas is compared with the code with $37.5\%$ zeros ( denoted as RZCOD ) and the code with $68.75\%$ zeros (denoted as SCOD) of same order in Fig. \ref{fig1} under peak power constraint.
Similarly, in Fig. \ref{fig2}, the performance comparison of the corresponding codes under average power constraint is shown. The average power constraint performance of NZCOD matches with that of the RZCOD and SCOD, while the NZCOD performs better than the other two codes under peak power constraint as seen in Fig. \ref{fig1}. Similarly, for $32$ antennas, the performance comparison shown in Fig. \ref{fig1_32} and in Fig. \ref{fig2_32}
establish the fact that the NZCOD performs better than the others under peak power constraint while under average power constraint, all the codes perform identically.

\section{Conclusion}
\label{sec4}
We have constructed square complex orthogonal designs for all power of $2$ antennas such that none of the entries in the matrix is zero. These codes have significant advantage over the existing codes in term of PAPR as the existing codes has zeros in its matrices. The only sacrifice that is made in the construction of these codes is that the signaling complexity of the these codes is marginally greater than the existing codes (with zero entries) as some of the entries in the codes of this paper consist of co-ordinate interleaved variables. An interesting direction to pursue is to investigate whether it is possible to construct codes with no zero-entry and also having lesser signaling complexity than the ones constructed in this paper. We conjecture that such codes do not exist.


\end{document}